\documentclass[draft]{agujournal2019}
\usepackage{url}
\usepackage{amsmath}
\usepackage{amsfonts}
\usepackage{booktabs}
\usepackage{soul}

\draftfalse

\begin{document}

\title{Conceptual framework for the application of deep neural networks to surface composition reconstruction from Mercury’s exospheric data}

\authors{Adrian Kazakov\affil{1}, Anna Milillo\affil{1}, Alessandro Mura\affil{1}, Stavro Ivanovski\affil{2}, Valeria Mangano\affil{1}, 
Alessandro Aronica\affil{1}, Elisabetta De Angelis\affil{1}, Pier Paolo Di Bartolomeo\affil{1}, Alessandro Brin\affil{1}, Luca Colasanti\affil{1}, 
Miguel Escalona-Morán\affil{3}, Francesco Lazzarotto\affil{4}, Stefano Massetti\affil{1}, Martina Moroni\affil{1}, 
Raffaella Noschese\affil{1}, Fabrizio Nuccilli\affil{1}, Stefano Orsini\affil{1}, Christina Plainaki\affil{5}, Rosanna Rispoli\affil{1},
Roberto Sordini\affil{1}, Mirko Stumpo\affil{1}, Nello Vertolli\affil{1}}

\affiliation{1}{INAF-IAPS, via Fosso del Cavaliere 100, 00133, Rome, Italy}
\affiliation{2}{INAF-Osservatorio Astronomico di Trieste, via Giambattista Tiepolo 11, 34143, Trieste, Italy}
\affiliation{3}{Augmented Intelligence Lab, Rua Lugo 2, 36470, Salceda de Caselas, Spain}
\affiliation{4}{INAF-Osservatorio Astronomico di Padova, Vicolo Osservatorio 5, 35122, Padova, Italy}
\affiliation{5}{ASI - Italian Space Agency, Via del Politecnico, 00133, Rome, Italy}

\correspondingauthor{Adrian Kazakov}{adrian.kazakov@inaf.it}

\begin{keypoints}
\item This study introduces the conceptual framework for the application of a machine learning model that predicts Mercury’s surface composition using simulated exospheric data.
\item The trained deep neural network is an estimator, a data-driven representation, of the surface-exosphere interactions.
\item This method has potential to enhance future studies of planetary exospheres, including Mercury’s, using data from space missions.
\end{keypoints}

\begin{abstract}
Surface information derived from exospheric measurements at planetary bodies complements surface 
mapping provided by dedicated imagers, offering critical insights into surface release processes, interactions 
within the planetary environment, space weathering, and planetary evolution. This study explores the feasibility 
of deriving Mercury’s regolith elemental composition from in-situ measurements of its neutral exosphere using 
deep neural networks (DNNs). We present a supervised feed-forward DNN architecture—a multilayer perceptron 
(MLP)—that, starting from exospheric densities and proton precipitation fluxes, predicts the chemical elements of 
the surface regolith below. It serves as an estimator for the surface-exosphere interaction and the processes 
leading to exosphere formation. Because the DNN requires a comprehensive exospheric dataset not available 
from previous missions, this study uses simulated exosphere components and simulated drivers. Extensive training 
and testing campaigns demonstrate the MLP’s ability to accurately predict and reconstruct surface composition 
maps from these simulated measurements. Although this initial version does not aim to reproduce Mercury’s actual 
surface composition, it provides a proof of concept, showcasing the algorithm’s robustness and capacity for 
handling complex datasets to create estimators for exospheric generation models. Moreover, our tests reveal 
substantial potential for further development, suggesting that this method could significantly enhance the analysis 
of complex surface-exosphere interactions and complement planetary exosphere models. This work anticipates 
applying the approach to data from the BepiColombo mission, specifically the SERENA package, whose nominal 
phase begins in 2027.
\end{abstract}

\section*{Plain Language Summary}
Mercury is subjected to a variety of environmental effects that influence the complex interactions between its outer planetary 
layers. This study focuses on Mercury’s interaction with its surrounding space environment, particularly the connection between 
its surface and its thin exosphere (a layer of gases around the planet). We developed a computer model that uses data from 
the exosphere to predict the chemical elements present on the surface. This model, a deep neural network, is trained on 
simulated data that includes gas densities in the exosphere and particles from the solar wind. By learning from this information, 
the model can estimate the processes that form Mercury’s exosphere, such as vaporization from micrometeoroid impacts and 
the release of gas from particle bombardment. Our tests show that this method is able to predict surface compositions, which 
could help scientists better understand the interactions between the planet's surface, its exosphere, and its environment. 
This research is especially relevant for future space missions, like the ESA-JAXA BepiColombo mission, which will 
begin collecting real data from Mercury in 2027. Our method shows further potential to enhance how scientists interpret that 
data and provide new insights into Mercury’s dynamics and evolution.

\section{Introduction}

Celestial bodies within our Solar System are continuously influenced by external forces such as solar wind, 
solar radiation, and micrometeoroids. These agents contribute to their reshaping by adding, removing, 
altering, or relocating material, affecting both their surfaces and atmospheres. Mercury's atmosphere, 
being exceptionally tenuous, is known as an exosphere - a planetary envelope where constituent particle 
collisions are so infrequent that their trajectories are essentially ballistic \cite{milillo1,domingue}.
This exosphere arises from a variety of environmental interactions with Mercury's surface. 
The external factors acting on the planet, such as dust particles, solar wind protons, and heavy ions, 
as well as solar radiation and intense heat, have profound effects on the exosphere \cite{killen1}.

The active processes in the formation of Mercury's tenuous atmosphere are widely discussed in 
the literature \cite{mura1,wurz3,killen2,Grava2021,Milillo2023}. Four predominant processes 
release atoms and molecules from the surface into the exosphere: micrometeoroid impact vaporization (MIV),  
sputtering after solar wind and heavy ion impacts (SP), thermal desorption (TD), and photon-stimulated 
desorption (PSD). MIV and SP are particularly indicative of the regolith composition below, as they 
involve higher energy transfers capable of dislodging neutral species from their minerals. Conversely, 
TD and PSD, being less energetic, tend to release atoms and molecules that are weakly bonded to minerals, 
such as volatile elements, most of which eventually fall back and are reabsorbed by the surface 
\cite{killen1,mura2,gamborino,Leblanc2023}. 
Once in the exosphere, the released particles undergo further transformations due to interactions 
with radiation pressure, photons, and charged particles. Such interactions can modify the charge, 
chemical state, and movement of these exospheric constituents. However, in a first approximation
in the sparse exosphere, the atomic and molecular abundances resulting from these actors could 
be traced back to the planet, connecting the surface properties, like composition, mineralogy, and physical 
state to the different processes and the dynamics of matter around the planet \cite{milillo3,rothery}. 
There has already been direct evidence that this is the case for the Magnesium exosphere, which is directly
related to the Magnesium-rich surface below, as shown by \citeA{merkel}.

To gain a deeper understanding of Mercury's exosphere, scientists use sophisticated models to simulate the 
active processes and their effects on the planetary surface, thereby attempting to replicate the generation of the 
exosphere. This extensively applied method compares the results of simulations to those measured from space 
(e.g. \citeA{sarantos}, \citeA{cassidy2}, \citeA{plainaki}) or from Earth (e.g. \citeA{wurz1}, \citeA{mura2}, 
\citeA{mangano2}). However, the inherent 
complexity of these interactions, which includes electromagnetic, chemical, mechanical, thermal, and other effects, 
and the validity of the chosen parameters in their mathematical representation add significant challenges. Some of 
these effects have not been precisely evaluated for each release process, leading to a broad range of simulated 
results with considerable uncertainty, depending on the assumptions made at the outset. To address these, a 
multifaceted approach is required, involving the refinement of models through improved measurements, continual 
reassessment of the model structure, and advanced statistical methods to better understand and quantify uncertainties.

In parallel, machine learning algorithms, particularly deep neural networks (DNNs), offer a novel approach to capture 
the relationships between the variables. These algorithms can approach, to some extent, the data 
generation mechanisms \cite{russell,lecun,goodfellow1}, providing a tool to explore in depth the relationships between 
the components of Mercury’s environment. This work represents a pioneering attempt to apply a machine learning 
algorithm to the study of Mercury's environment. The primary objective of this study is not to reproduce the actual Mercury 
environment but to validate the methodological framework—that is, to demonstrate that a multilayer perceptron (MLP) can 
reconstruct an underlying compositional map from exospheric data, which, in this case, has been substituted with simulated data. 
We have intentionally chosen a range of simplified or not-yet-fully-realistic composition models to prevent the machine learning 
pipeline from becoming biased by an already well-constrained or overly specific map of Mercury’s surface.
In fact, we aim to demonstrate how DNNs, particularly multilayer perceptrons (MLPs), 
can be employed within the data analysis of Mercury's exosphere to reconstruct the elemental surface map underneath. Suitable for 
nonlinear regression tasks, DNNs scale effectively with increasing training data and input parameters \cite{minsky,hinton1,ciresan}, 
often yielding improved performance when appropriately structured and tuned, offering a promising direction for tackling the 
complexities inherent in modeling Mercury’s exosphere.

This study builds upon the preliminary work of \citeA{kazakov} by extending, refining, and further advancing 
the application of deep neural networks in predicting Mercury's surface composition from exospheric measurements. 
It involves predicting the elemental composition of the surface using data from more sophisticated and realistic models 
that simulate all major processes—MIV, SP, TD, and PSD—and consider the influence of solar radiation pressure and 
photolysis on the exosphere. The multilayer perceptrons have been extensively optimized through a comprehensive 
exploration of their building blocks, resulting in the development of a robust predictive algorithm.

In \textbf{Section 2}, we introduce and detail the algorithm - the multilayer perceptron deep neural network - outlining 
its structure for the multivariate regression task of predicting surface composition. This section methodically breaks down 
each component of the algorithm, providing a comprehensive guide for constructing effective neural network architectures. 
\textbf{Section 3} delves into the mechanisms behind exospheric data generation, encompassing the models of Mercury's 
surface, its environment, and the processes generating the exosphere. It also elaborates on the creation of the datasets 
used in the algorithms, including feature selection and data augmentation for the DNN inputs. The findings from an extensive 
training and testing campaign are explored in \textbf{Section 4}. This section details the selection of the architectures' 
hyperparameters, offering insights into the optimal choices within the DNN hyperparameter space to develop an effective 
MLP DNN. The testing of the algorithms is presented, showcasing their performance on a variety of surface-exosphere pairs 
and culminating in the visual demonstration of reconstructed surface elemental composition maps. \textbf{Section 5} discusses
the assumptions and implications of the current method.
Finally, the paper concludes in 
\textbf{Section 6}, outlining future perspectives for the method, highlighting its potential and the wide scope for further research and development in this field.

\section{Method}

\subsection{Prediction Task and General Characteristics of the Method}

\begin{figure}[h!]
		\includegraphics[scale=0.6]{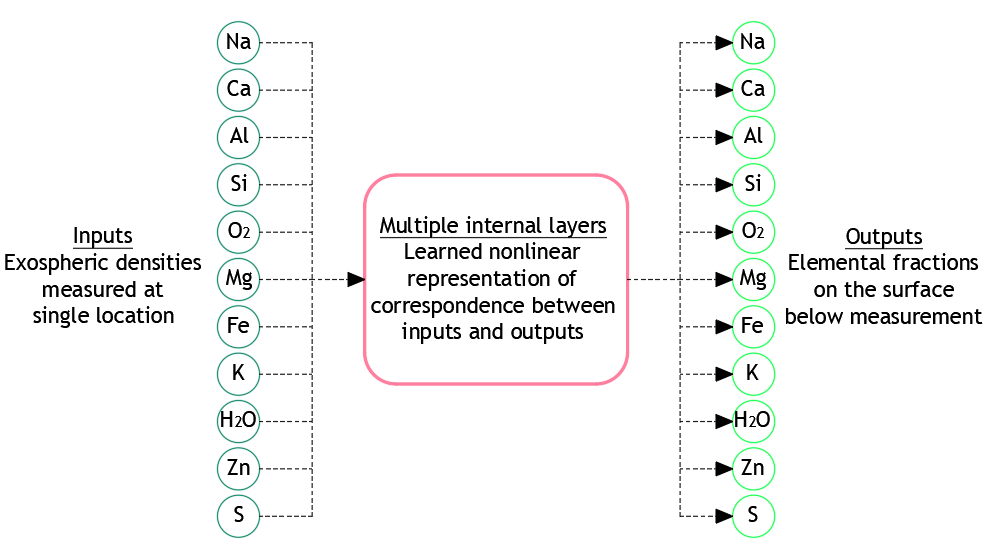}
		\centering
		\caption{DNN prediction task schematics. The input to the neural network on the left are the exospheric densities 
		at a single location in the exosphere.
		The output is the relative surface elemental composition as fractions summing up to 1 at a surface area just below
		the exospheric measurement.
		The hidden layer box consists of multiple layers and represents the complex, often nonlinear, 
		relationships between the inputs and the outputs of the neural net.}
  		\label{fig:MLP-pred-task}
	\end{figure}

	In this study, we develop a machine learning algorithm and apply it to supervised 
	multivariate regression of exospheric data at Mercury using a multilayer perceptron deep neural 
	network. The objective of this DNN is to infer the regolith source material, believed to be 
	a primary contributor to Mercury's exosphere. Specifically, the DNN predicts elemental surface composition 
	fractions from exospheric density measurements, governed by the equation:

	\begin{equation}
		\sum_{i=1}^{n} \hat{y_{i}} = 1,
	\end{equation}

	where $\hat{y_{i}}$ is the fraction of an elemental species predicted by the neural network to be present in 
	the surface area below the exospheric measurement, and $n$ is the total number of elements in the prediction task
	(Figure \ref{fig:MLP-pred-task}).

	The methodological and algorithmic developments in this study include:
	\begin{itemize}
		\item Laying the foundation for the algorithm's application for predicting surface elemental composition.
		\item Building DNNs capable of operating in a multi-process environment, integrating the four primary 
		active processes (MIV, SP, TD, PSD) responsible for neutral species release.
		\item Implementing a data production model based on plausible assumptions for the exosphere generation processes.
		\item Training the algorithms using augmented datasets to enhance performance.
		\item Employing hyperparameter tuning to optimize the DNN design parameters.
		\item Presenting a basis for the exploration of the physical processes parameter space.
	\end{itemize}

	Ultimately, the goal of the MLP DNN is to encapsulate the complex relationships between various surface processes 
	and their impact on the generation of the exosphere, thereby formulating an estimator for these interactions.

	In this foundational proof-of-concept paper, we aim to validate the proposed methodological framework. Specifically, we explore 
	the capability of a multilayer perceptron to infer an underlying compositional map from simulated exospheric data, derived under generalized 
	yet realistic conditions, and reverse-engineer the processes assumed in the data generation model. The diverse paths 
	for development of the method beyond this initial validation—including its adaptation to real data and its application to alternative 
	analytical and data-driven physical representations—are expanded in Section 6.

\subsection{Deep Neural Network Architecture}

\begin{figure}[h!]
		\includegraphics[width=\linewidth]{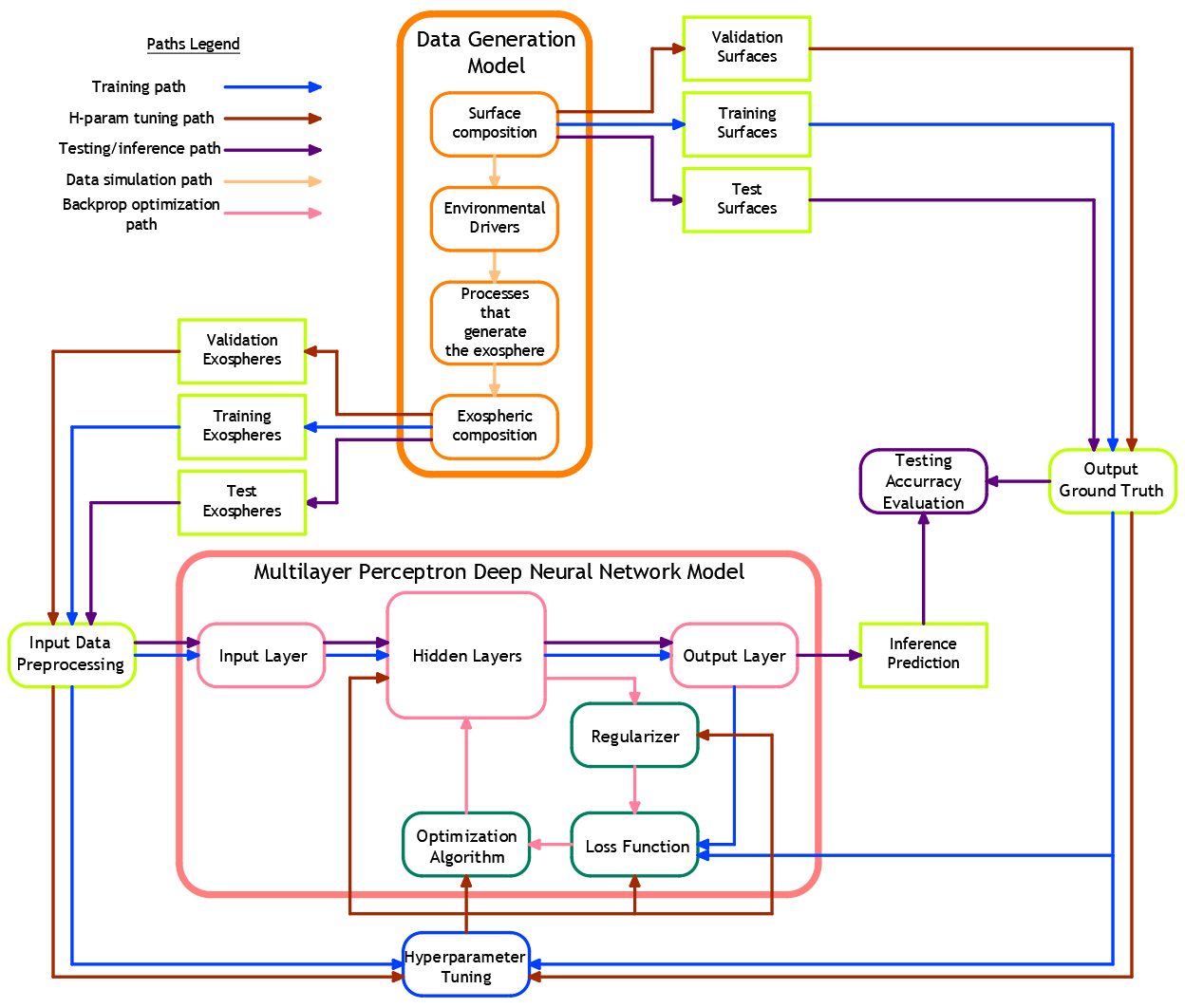}
		\centering
		\caption{MLP DNN architecture overview. The data generation model produces both the inputs and the outputs
		for training, validating, and testing the algorithm. This simulated data is passed through the MLP DNN model in the training, 
		hyperparameter tuning and testing phases, respectively. The backpropagation optimization uses the loss function, 
		regularizer and optimization algorithm to adjust the weights (internal parameters) of the neural network. In a 
		separate process, the hyperparameter tuner adjusts/optimizes the MLP DNN by minimizing the errors on the 
		validation dataset. After the final training, the previously unseen data from the testing sets is passed through 
		the network and the accuracy of the predictions (performance of the network) is evaluated. Note: This proof-of-concept
		study trains and tests the DNN's capability to estimate the processes that are simulated via the Data Generation Model. 
		As one of the goals of future studies will be to construct a data-driven representation of real physical 
		processes, one approach in that case would be to replace the Data Generation Model with the real physical data generation mechanisms.
		Other alternative approaches are discussed in Section 6.}
  		\label{fig:MLP-arch}
	\end{figure}

\begin{figure}[h!]
		\includegraphics[width=\linewidth]{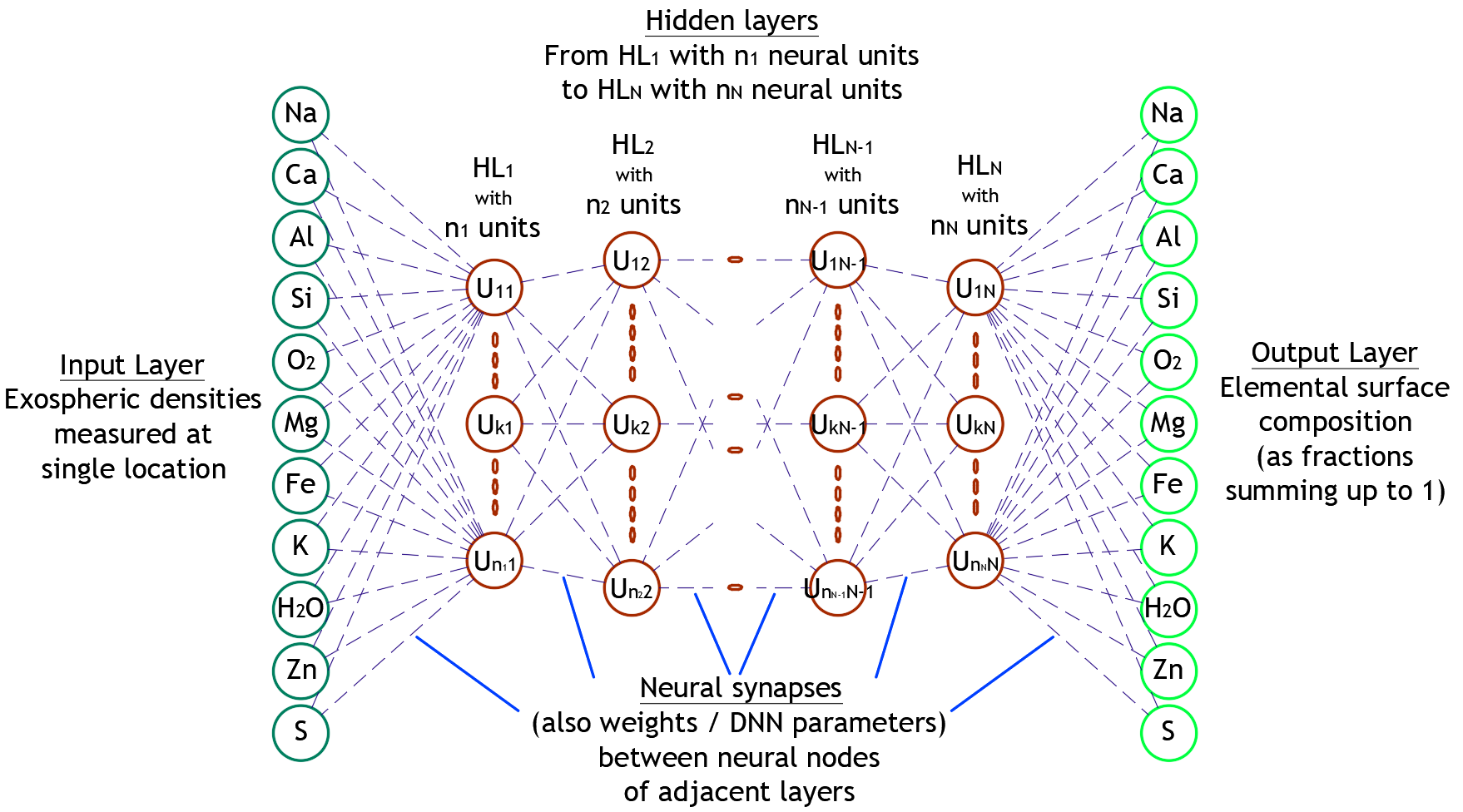}
		\centering
		\caption{MLP DNN basic structure schematics. Exospheric densities form the input layer on the left.
		The output layer is formed from the relative surface elemental composition as fractions summing up to 1.
		There are $N$ number of hidden layers with $n_1$ to $n_N$ number of neural units (neurons). This structure
		represents the relationships between the inputs and the outputs of the MLP. The neural synapses, connections between
		the neural units, form the weight (DNN parameter) matrices $W_1$ to $W_{N+1}$.}
		\label{fig:MLP-est-basic}
	\end{figure}

	Our DNN follows a standard multilayer perceptron architecture that enables complex data processing through a structured 
	network of layers: an input layer, multiple hidden layers for nonlinear transformations, and an output layer 
	for predictions. The network's effectiveness hinges on key components like the loss function, which guides 
	accuracy improvements, and the regularizer, which ensures generalizability. Efficiently chosen optimization 
	algorithms and precise hyperparameter tuning further enhance the network's performance. Figure 
	\ref{fig:MLP-arch} illustrates this interplay, crucial for tasks like analyzing Mercury's exosphere. The inner connectiveness 
	of the MLP DNN neural units is shown on Figure \ref{fig:MLP-est-basic}. The following paragraphs give an overview
	of each DNN component, while further details, including derivations, equations, and expanded discussion on their predictive
	capabilities, can be found in \textbf{Appendix A}.

	\subparagraph{Input, Hidden, and Output Layers}\hfill

		The input layer of the MLP introduces data, in our case Mercury’s virtual exospheric density measurements, 
		into the network, with each neuron representing a distinct data feature. For example, distinct elemental 
		density measurements are represented by separate neurons. The inputs are normalized to zero mean 
		and unit variance - a preprocessing step aiming to prevent large or unbalanced feature scales from biasing the 
		training \cite{goodfellow1}.

		Multiple fully connected hidden layers with a substantial number of neurons allow the MLP to capture 
		the nuances of Mercury's exospheric composition and the underlying processes that govern it. The ReLU 
		activation ($\max(0, z)$) is chosen for its computational simplicity and effectiveness in deeper networks \cite{glorot}.

		The output layer in our DNN corresponds to elemental composition fractions for each element 
		in our prediction task. We employ the softmax activation function to convert raw network outputs into 
		valid fractions that sum to 1, reflecting physical constraints on Mercury’s surface composition.
		This configuration allows the network to deliver accurate, meaningful predictions of 
		Mercury’s surface composition, synthesizing the representation insights gained from all previous layers.

	\subparagraph{Loss Function, Regularization, and Optimization}\hfill

		The loss function plays a pivotal role in guiding the optimization process, quantifying the discrepancy 
		between the network's predictions and the actual target values to gauge model accuracy. In our most 
		successful MLP tests on predicting Mercury's surface elemental composition, the Kullback-Leibler (KL) 
		divergence \cite{cover} has proven particularly effective. It measures how one probability distribution, 
		representing the predicted elemental composition (the output from the MLP), diverges from the actual 
		distribution (the true elemental composition). The KL divergence was preferred for our regression task because 
		it aligns better with the probabilistic requirements, focusing on relative proportions rather than absolute quantities 
		of elements.

		We apply L2 regularization (weight decay) on the weights of each hidden layer \cite{bishop2}. This technique 
		constrains the magnitude of the weights, preventing them from becoming excessively large and helping to avoid 
		overfitting the model to the specific dataset used for training. In a multivariate regression task such as ours, where the 
		model needs to understand complex relationships between various features in the surface-exosphere interaction at Mercury, 
		L2 regularization helps in maintaining a balance between MLP model complexity and its ability to generalize. The addition 
		of this regularization term (penalty) to the loss function thus ensures that the model not only fits the training data 
		well but also maintains the flexibility to perform accurately on new, unseen data.

		The training of our multilayer perceptron for predicting Mercury's surface composition employs the Adam 
		optimization algorithm, a refinement of stochastic gradient descent known for its effectiveness with 
		large-scale data and complex models \cite{kingma}. Through backpropagation \cite{rumelhart1, rumelhart2}, 
		the weights of the combination matrices for each hidden layer are adjusted following their gradients with respect to the 
		loss function. The stochastic nature of the gradient descent implies that learning iterations are not performed on the 
		entire dataset but rather on a random subset known as a mini-batch.

	\subparagraph{Hyperparameter Tuning}\hfill

		Key hyperparameters of our MLP DNN, such as the learning rate, the number of hidden layers, the mini-batch size, 
		and the L2 coefficient, were systematically optimized using a Bayesian approach with Gaussian Processes \cite{bergstra}. 
		The tuning process was facilitated by the scikit-optimize library \cite{scikit-opt}, which utilizes a prior probability distribution 
		function to identify the hyperparameter configuration that minimizes the total loss on a hold-out validation dataset. This 
		systematic adjustment of hyperparameters not only enhances learning capabilities and overall performance, but also optimizes 
		the balance between model complexity and efficiency.

	\subparagraph{Performance Metrics}\hfill

		To evaluate the performance of our machine learning model, we utilize both customized and standard metrics 
		to ensure precise and insightful quantitative assessments. Our primary metric, the Euclidean 
		similarity 4 (ES4), integrates elements of Euclidean distance and cosine similarity, providing a nuanced measure 
		of prediction accuracy by considering both magnitude and directionality in multidimensional space:
	
		\begin{equation}
			\textrm{ES4} = \left(1 - \frac{\sqrt{\sum_{i}(\boldsymbol{\hat{y}_i} - \boldsymbol{y_i})^2}}{\sqrt{\sum_{i}\boldsymbol{y_i}^2}}\right) \times \left(\frac{\boldsymbol{\hat{y}_i} \cdot \boldsymbol{y_i}}{\|\boldsymbol{\hat{y}_i}\| \|\boldsymbol{y_i}\|}\right),
		\end{equation}
	
		where $\boldsymbol{\hat{y_i}}$ and $\boldsymbol{y_i}$ represent the predicted and actual surface compositions, 
		respectively.

		Additionally, we evaluate the model using the R-squared ($R^2$) metric and residuals, both absolute and relative,
		which provide further granularity in understanding the model's performance. By combining these metrics, we achieve a 
		multidimensional evaluation of our DNN's performance, encompassing both the accuracy of individual predictions and the 
		model's overall ability to capture the complexity of the data. This comprehensive assessment not only ensures validation 
		of the model's outputs but also sheds light on areas for potential improvement, thereby contributing to the refinement of 
		the model's predictive capabilities.

\section{Data Generation Model}

For the development of our MLP DNN, we employ a data generation model consisting of three components: 
a surface composition model, an exosphere generation model, and a model for the distributions of the drivers 
of surface release processes. The surface model in our study is treated as a set of nearly 
unconstrained composition maps for each species. These maps are what the DNN aims to reconstruct. 
The exosphere model captures all the main physical processes that link the surface, external environment 
conditions, and the exosphere. The drivers model comprises maps of the surface where external drivers, 
such as ions, photons, and micrometeoroids, are active. These maps are derived from simplified functions 
in this prototype network. In the following sections, we describe the models used for these three components 
in greater detail.

Our focus extends beyond the inherent complexities of Mercury's exosphere to include a thorough description 
of the physical parameter space representing the processes behind exospheric formation. 
This parameter space is an important element in our simulation approach, allowing us to explore and represent a variety of 
planetary environmental conditions. We consider a broad set of physical processes relevant to the surface-exosphere
interaction, spread across a highly multidimensional parameter space that encompasses both variable and
fixed processes and drivers. This space can be further subdivided into two parts: the parameter subspace
constructed from physical interactions that are explicitly or implicitly included in our model, and a
secondary parameter space describing aspects omitted due to limited computational resources or incomplete
scientific understanding.

The three components of the data generation model comprehensively describe the physical processes shaping the 
surface-exosphere interaction, allowing us to define a particular region in the parameter space. This region 
governs the data distribution that our multilayer perceptron deep neural network aims to estimate. 
The objective of this work is to demonstrate that our trained MLP DNN can closely approximate the most representative 
region in the physical parameter space and serve as a robust estimator of the data generation mechanism. 
Understanding this concept is crucial for grasping how our DNN represents the
surface-environment-exosphere relationships in a data-driven approach, laying the foundation for further exploration 
of both the parameter space itself, and the capability of the algorithm to approach different regions in it.

\subsection{Surface and Regolith}

	Even if the Mercury surface composition has been partially identified by MESSENGER data \cite{VanderKaaden2017,Nittler2019}, 
	for the purposes of this study—specifically, to validate our method—we decided to consider surface compositions 
	that are in a broader range around the composition ratios expected for the main mineralogical components of Mercury.
	We selected a specific set of minerals (Table \ref{tab:miner_base}) believed to be present on 
	Mercury’s surface \cite{wurz3}. While this proposed mineralogy may not fully represent Mercury’s derived composition 
	post-MESSENGER, our approach preserves the method’s adaptability, ensuring that the chosen compositions remain hypothetical 
	rather than strictly aligned with Mercury’s known mineralogy.

	These minerals are assumed to exist in varying proportions, contributing to the overall mineral composition of the regolith. 
	These proportions delineate zones characterized by dominant 
	primary minerals and their secondary counterparts, enforcing the presence of some of the minerals on the 
	surface. Additionally, constraints on the minimal fractions of specific minerals and the presence of water ice 
	further refine this parameter subspace.

	\begin{table}[h]
	    \centering
		\caption{Mineral composition considered in the baseline surface model. The minerals are decomposed 
		via the classical additive method to elemental species. The decomposition captures some of the 
		relationships between mineralogy and elemental composition, while others are omitted (e.g. decomposition 
		of water ice, or decomposition to heavier molecules). The mean mineral fraction reported in this table
		is for all the datasets generated in this study - 204,768 surface tiles.}
		\resizebox{\textwidth}{!}{%
	    \begin{tabular}{l l l l c c}
	        \toprule
	        \textbf{Mineral Name} & \textbf{Chemical Formula} & \textbf{Decomposed to} & \textbf{Rarity} & \textbf{Mean Fraction} & \textbf{Range} \\
	        \midrule
	        Anorthite    & CaAl$_2$Si$_2$O$_8$  & Ca, 2Al, 2Si, 4O$_2$  & -    & 0.134 & 0.049 - 0.319 \\
	        Albite       & NaAlSi$_3$O$_8$      & Na, Al, 3Si, 4O$_2$   & -    & 0.140 & 0.051 - 0.341 \\
	        Orthoclase   & KAlSi$_3$O$_8$       & K, Al, 3Si, 4O$_2$    & -    & 0.134 & 0.050 - 0.313 \\
	        Enstatite    & Mg$_2$Si$_2$O$_6$    & 2Mg, 2Si, 3O$_2$      & -    & 0.137 & 0.053 - 0.312 \\
	        Diopside     & MgCaSi$_2$O$_6$      & Mg, Ca, 2Si, 3O$_2$   & -    & 0.141 & 0.053 - 0.336 \\
	        Ferrosilite  & Fe$_2$Si$_2$O$_6$    & 2Fe, 2Si, 3O$_2$      & -    & 0.137 & 0.052 - 0.329 \\
	        Hedenbergite & FeCaSi$_2$O$_6$      & Fe, Ca, 2Si, 3O$_2$   & Rare & 0.065 & 0.010 - 0.213 \\
	        Sphalerite   & ZnS                  & Zn, S                & Rare & 0.069 & 0.012 - 0.243 \\
	        Water Ice    & H$_2$O               & H$_2$O                & Rare & 0.044 & 0.012 - 0.136 \\
	        \bottomrule
	    \end{tabular}
		}
		\label{tab:miner_base}
	\end{table}

	Among the included minerals, hedenbergite (primarily deposited from meteorites), sphalerite (resulting from 
	volcanic activity), and water ice are considered as rare minerals. The inclusion of these species, even though unlikely to be
	significant contributors on Mercury, aims at exploring the method's capability to recognize their specifics and improve its generalization. 
	Nevertheless, their presence is strongly reduced in the 
	random surface generation, compared to the other six minerals. Furthermore, in the randomized creation of 
	the surface, they are not allowed to be distributed everywhere on the surface. The overall minerals
	used in our datasets, including the split to their constituent elemental species (atoms or molecules), 
	are reported in Table \ref{tab:miner_base}. This implies an assumption that the surface, on average, 
	encompasses a complete pool of atoms and molecules derived by stoichiometric (atomic) decomposition from these minerals, 
	which are then subjected to external environmental forces. This approach, as an approximation, considers the full fraction of volatile 
	species (such as Na, K, H$_2$O, S, and O$_2$) as being readily available for release into the exosphere as they 
	are loosely bound to the regolith grains. The list of elements resulting from the mineral break down is presented 
	in Table \ref{tab:elem_base}. The elemental composition resulting from this process represents the ’actual’ or 
	’ground truth’ data that we compare with the predictions obtained by our algorithms.

	\begin{table}[h]
		\centering
		\caption{Elemental composition considered in the baseline surface model. The elements are broken down from 
		minerals in the classical additive stoichiometric method. The ranges of variation for each element are in the 
		last column.}
		\resizebox{\textwidth}{!}{%
		\begin{tabular}{l l l l c c}
	        \toprule
	        \textbf{Element Name} & \textbf{Designation} & \textbf{From Mineral} & \textbf{Rarity} & \textbf{Mean Fraction} & \textbf{Range} \\
	        \midrule
	        Aluminium     & Al   & Anorthite, Albite, Orthoclase          & -         & 0.072 & 0.043 - 0.113 \\
	        Calcium       & Ca   & Anorthite, Diopside, Hedenbergite      & -         & 0.047 & 0.025 - 0.072 \\
	        Iron          & Fe   & Ferrosilite, Hedenbergite              & -         & 0.048 & 0.020 - 0.098 \\
	        Sodium        & Na   & Albite                                 & -         & 0.019 & 0.007 - 0.042 \\
	        Oxygen        & O$_2$ & All, except Sphalerite, Water Ice    & Dominant  & 0.423 & 0.391 - 0.435 \\
	        Sulfur        & S    & Sphalerite                             & Rare      & 0.012 & 0.002 - 0.047 \\
	        Water Vapor   & H$_2$O & Water Ice                          & Rare      & 0.008 & 0.002 - 0.026 \\
	        Zinc          & Zn   & Sphalerite                             & Rare      & 0.012 & 0.002 - 0.047 \\
	        Silicium      & Si   & All, except Sphalerite, Water Ice     & Dominant  & 0.282 & 0.258 - 0.301 \\
	        Potassium     & K    & Orthoclase                             & -         & 0.018 & 0.007 - 0.041 \\
	        Magnesium     & Mg   & Enstatite, Diopside                    & -         & 0.059 & 0.030 - 0.109 \\
	        \bottomrule
		\end{tabular}
		}
  		\label{tab:elem_base}
	\end{table}

	For our modeling, we need to consider not only the composition map but also some characteristics of 
	the planetary surface and regolith (the loose, heterogeneous material covering solid rock), which are part of 
	our simulation model. These include the influence of surface composition, texture, and physical, chemical, and 
	thermal properties, all in the context of forming the modeling parameter space and defining a region within 
	that space.

	Firstly, the surface in our model is represented as a grid comprised of 36$\times$18 surface tiles in a
	modified Mercator projection. Each tile measures 10$^{\circ}$$\times$10$^{\circ}$, which, at the equator, 
	translates to approximately 425 km$\times$425 km. This averaging of composition inevitably reduces the 
	complexity of the parameter space, as it results in the loss of finer details in the spatial relations of the
	spread of the different species - elemental and mineral - on the planetary surface. However, it is important to note 
	that the potential resolution of surface composition maps reconstructed from exospheric measurements by an 
	orbiting spacecraft cannot be significantly higher. 
	
	Conversely, while our model omits certain surface qualities such as grain sizes, slope angles, and roughness, 
	we do incorporate a simplified representation of porosity and the presence of microshadows in the ion-sputtering 
	process acting on the surface.

\subsection{Environmental Drivers}

	In our model definition, we incorporate the environmental conditions and various factors that contribute 
	to changes in the sources or processes for the release of material from the planetary surface into the exosphere. 
	These drivers encompass solar radiation, dust particles, and charged particles that enable surface material to 
	escape into the exosphere. Mercury's proximity to the Sun significantly influences its interaction with the 
	surrounding environment. Its highly eccentric orbit, varying between 0.31 and 0.46 astronomical units (AU), 
	causes external conditions such as thermal radiation, photon flux, and solar wind intensity to fluctuate according 
	to its distance from the Sun. In our model, we specifically focus on conditions at perihelion. This close distance 
	markedly impacts the intensity of the solar influence, thereby affecting the range of effects and processes 
	contributing to the generation of the exosphere. 

	One such influence is the equivalent photon flux, which is the photon flux at Earth's orbit adjusted  
	for Mercury's closer position to the Sun by a factor of $1/r^2$, where $r$ is the distance to the Sun in AU. 
	A photon flux at Earth of $3.0\times10^{15}$ cm$^{-2}$s$^{-1}$ is considered \cite{mura2} resulting
	in an equivalent photon flux of $3.1\times10^{16}$ cm$^{-2}$s$^{-1}$.

	Another environmental aspect is the activity level of the Sun, which we have assumed to be at a moderate level, 
	devoid of extreme events such as coronal mass ejections or solar flares. This assumption sets the conditions 
	for a solar wind velocity of 450 km/s and a solar wind density of 60 cm$^{-3}$ at Mercury's perihelion 
	\cite{Wilson2018}.

	Furthermore, the dust environment around Mercury is considered for particles smaller than 100 $\mu$m in 
	diameter, with a mean flux of $1.0\textrm{x}10^{-16}$ g/cm$^{2}$s and mean velocity of 20 km/s
	in Mercury's vicinity, in agreement with the modal impact velocity reported by \citeA{cintala}. This is compared to a 
	planet velocity at perihelion of 59 km/s. However, our model does not differentiate between the origins of these 
	dust particles—whether they come 
	from the Main Belt Asteroids, Jupiter Family Comets, Oort Cloud Comets, or Halley Type Comets—nor does 
	it consider the full ranges and exact distributions of particle sizes and velocities as in \citeA{Pokorny2018}.
	Additionally, no large meteorite impacts or increases of fluxes due to particularly dense cometary streams, 
	such as from comet Encke \cite{plainaki}, are considered. Grain size distribution influence of the dust particles 
	is also not represented in our physical parameter space.

\begin{figure}[h!]
		\includegraphics[width=\linewidth]{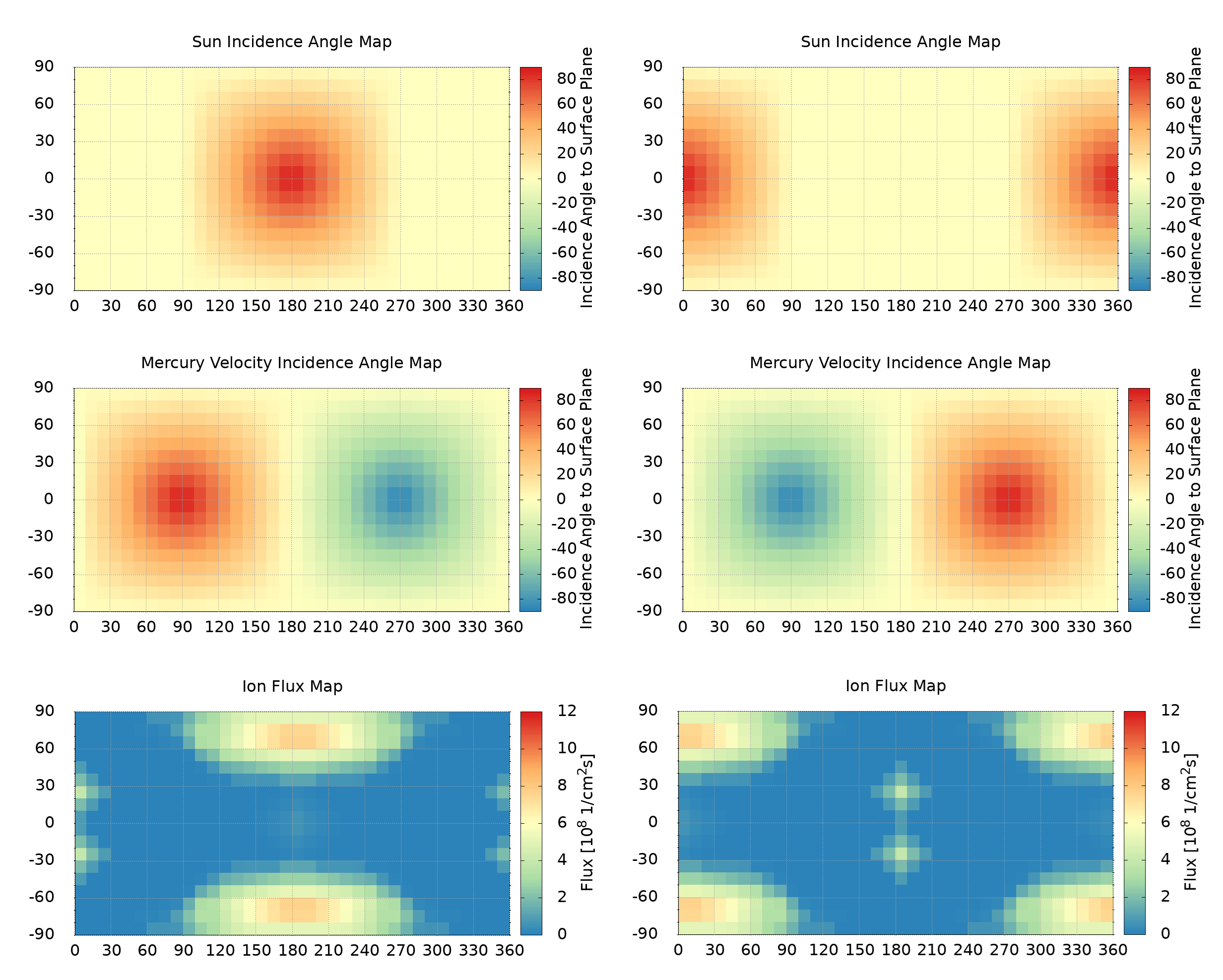}
		\centering
		\caption{Maps of incidence angles due to planet orientation at two consecutive perihelia. Perihelion 1 on the left
		and perihelion 2 on the right. The top panels represent the incidence maps of solar photons on the surface. The
		middle panels represent the angles between the planet's velocity and the surface normal. The bottom panels
		show the magnetic field cusp footprints on the surface. A shift of 180 degrees in longitude between the two
		perihelia is notable due to the spin-orbit resonance of Mercury.}
  		\label{fig:planet_orient}
	\end{figure}

	The environmental conditions on the planet itself present a diverse range of parameters due to varying 
	exposure to sunlight and shadow, as well as differences in particle fluxes on the planet's leading and trailing 
	sides due to its high orbital velocity. Our model incorporates detailed maps that illustrate solar incidence 
	angles and planetary velocity incidence angles at Mercury’s perihelion (Figure \ref{fig:planet_orient}). It is crucial 
	to recognize Mercury’s unique orbit-spin resonance, which alternates the sides facing the Sun at the same 
	true anomaly angle in successive orbits, a fact that we have taken advantage of later in our study.

	Our study does not encompass the full spectrum of influences that the interplanetary magnetic field and Mercury's 
	own magnetic field might have on the planet's surface and exosphere. The sole aspect of Mercury's magnetic field 
	that our model takes into account is the presence of open magnetic field lines. These lines serve as conduits for 
	charged particles, channeling them through magnetic field cusps directly onto the planet's surface at specific locations 
	known as cusp footprints \cite{Massetti2003}. The impact of this process is significant, as it concentrates ion 
	bombardment in particular areas, altering the surface composition and influencing the generation of the exosphere. 
	We have delineated the shape and relative impact areas of these cusp footprints in Figure \ref{fig:planet_orient}, 
	highlighting the regions on Mercury's surface that are most affected by the ion funneling. In this initial version 
	of our DNN, we assume a centered dipole, which leads to symmetric cusp footprints in both hemispheres. These zones 
	can also change dynamically, often over very short timescales. In our study, we have considered a fairly wide zones of 
	ion precipitation. A more dynamic dependence on the magnetic reconnection rate driven by the Interplanetary Magnetic 
	Field (IMF) strength and orientation is neglected. 

	A summary of the environmental conditions considered in our study and their contribution further to the release
	processes and to the represented parameter space are given in Table \ref{tab:env_base}.

	\begin{table}[h]
	    \centering
		\caption{Environmental parameters and conditions and their effects on the populations of particles or
		other processes that release particles from the surface.}
		\resizebox{\textwidth}{!}{%
	    \begin{tabular}{l l l l l l}
	        \toprule
	        \textbf{Group} & \textbf{Condition} & \textbf{Units} & \textbf{Value} & \textbf{Sources Affected} & \textbf{Processes Affected} \\
	        \midrule
	        \textbf{Star Activity} & Solar wind velocity & km/s & 450 & Proton flux & SP \\
	                              & Solar wind density  & 1/cm$^3$ & 60 & Proton flux & SP \\
	                              & Photon flux at Earth & 1/cm$^2$s & $3.0 \times 10^{15}$ & Equivalent photon flux & PSD \\
	        \midrule
	        \textbf{Comets and Asteroids} & Dust particle size & $\mu$m & 0-100 & Micrometeorite flux & MIV \\
	                                     & Mean flux of dust particles & g/cm$^2$s & $1.0 \times 10^{-16}$ & Micrometeorite flux & MIV \\
	                                     & Mean velocity of dust particles & km/s & 20 & Micrometeorite flux & MIV \\
	        \midrule
	        \textbf{Magnetic Fields} & Cusp footprints size & \multicolumn{2}{c}{Map} & Ion precipitation zone & SP \\
	                                & Cusp footprints location & \multicolumn{2}{c}{Map} & Ion precipitation zone & SP \\
	                                & Cusp footprints ion flux distribution & \multicolumn{2}{c}{Map} & Ion flux & SP \\
	                                & Cusp footprints area coefficient & - & 0.4 & Ion flux & SP \\
	        \midrule
	        \textbf{Planet} & Planet velocity & km/s & 59 & Micrometeorite flux & MIV \\
	                        & Planet orientation & deg & 0 (local solar time offset) & All sources influence zones & All \\
	                        & Distance from Sun & AU & 0.31 & All & All \\
	        \bottomrule
	    \end{tabular}
		}
		\label{tab:env_base}
	\end{table}

\subsection{Exosphere Generation Model}

	The exospheric model is used for generating our own simulated datasets for training and testing the DNN. 
	Our method is designed to be adaptable, not limited to a single model, but capable of reconstructing various 
	exosphere generation processes and predicting surface compositions. For this purpose, we decided 
	to use the \citeA{mura1} model, as it is a comprehensive exosphere generation model able to describe all 
	the main surface release processes for different species on Mercury as a function of external environmental 
	drivers, with the flexibility to tune the simulated exosphere with many relevant parameters and driver inputs. 
	It is important to note that the exosphere is non-collisional, meaning each released species does not interact 
	with others. Therefore, we can consider the exosphere as the sum of different exospheres for each species.

	An outline of the surface release processes used in our DNN is provided below, with more details available
	in \citeA{mura1,mura2}. These active processes include the four main ones, namely the micrometeorite 
	impact vaporization, sputtering from protons and heavy ions, thermal desoption, and 
	photon-stimulated desorption. Their respective sources, or drivers, are micrometeoroid fluxes, 
	precipitating ions through the open field lines of Mercury’s magnetic field, temperature effects on 
	the surface, and solar photons that impact the dayside surface. However, there are still significant 
	gaps in understanding these processes, making the problem not fully constrained in terms of 
	what is observed versus what the model can reproduce. Despite these uncertainties, we have chosen a 
	realistic range for their values.

	\subsubsection{Temperature map and Thermal desorption (TD)}

		TD becomes notably efficient at temperatures above 400 K \cite{mura1}. We consider this process 
		for the release of Na, K, H$_2$O, and S which are loosely bound to Mercury’s surface.

		We assume the subsolar point temperature on Mercury reaches 700 K at perihelion, while
		the night side registers a much lower temperature of 110 K. The temperature distribution across
		the surface adheres to a quarter-power law, ranging from a minimum of 110 K to a maximum
		of 700 K at subsolar point:

		\begin{equation}
			T_s(\phi,\theta) = T_{\textrm{min}} + (T_{\textrm{max}} - T_{\textrm{min}})(\textrm{cos}{\phi}\textrm{cos}{\theta})^{1/4},
		\end{equation}

		where $\phi$ represents the latitude and $\theta$ the longitude, as outlined in \citeA{mura1}.

		TD is considered only as a direct thermal ejection of species from the surface i.e. sticking coefficient 
		equal to 1. The flux of atoms or molecules resulting from TD is given by \citeA{mura1}:

		\begin{equation}
			\Phi_n = \nu Nc e^{\left(-\frac{U_d}{k_B T}\right)},
		\end{equation}

		where $\nu$ denotes the vibrational frequency of the species, $N$ the surface density of the regolith, $c$ the 
		fractional presence of the species within the regolith, $U_d$ the species' binding energy, $k_B$ the Boltzmann 
		constant, and $T$ the temperature at which desorption occurs. In this study, we have
		considered an invariable vibrational frequency for the relevant species, and the variance in their TD release flux 
		to be primarily due to differences in their binding energy and the surface composition.

		\begin{sidewaystable}
			\centering
			\caption{Parameters used in the processes of the exospheric model.}
			\begin{tabular}{l l l c c c c c c c c c c c}
	        \toprule
	        \textbf{Parameter} & \textbf{Process} & \textbf{Units} & \textbf{Al} & \textbf{Ca} & \textbf{Mg} & \textbf{Na} & \textbf{K} & \textbf{Fe} & \textbf{Si} & \textbf{Zn} & \textbf{S} & \textbf{O$_2$} & \textbf{H$_2$O} \\
	        \midrule
	        Dayside temperature & TD & K & \multicolumn{11}{c}{700} \\
	        Nightside temperature & TD & K & \multicolumn{11}{c}{110} \\
	        Surface density & TD & 1/cm$^2$ & \multicolumn{11}{c}{$7.5 \times 10^{14}$} \\
	        Vibrational frequency & TD & 1/s & \multicolumn{11}{c}{$1.0 \times 10^{13}$} \\
	        Binding energy & TD, SP & eV & 3.36 & 2.1 & 1.54 & 2 & 0.93 & 4.34 & 4.7 & 1.35 & 2.88 & 2 & 0.5 \\
	        \midrule
	        Mean photon flux & PSD & 1/cm$^2$s & \multicolumn{11}{c}{$3.1 \times 10^{16}$} \\
	        Beta coefficient & PSD & - & - & - & - & 1 & 1 & - & - & - & 1 & - & 1 \\
	        Threshold energy & PSD & eV & - & - & - & 0.052 & 0.02 & - & - & - & 0.06 & - & 0.01 \\
	        PSD cross section & PSD & 1/m$^2$ & - & - & - & \multicolumn{2}{c}{$1 \times 10^{-25}$} & - & - & - & $1 \times 10^{-25}$ & - & $1 \times 10^{-22}$ \\
	        \midrule
	        Mean flux of dust particles & MIV & g/cm$^2$s & \multicolumn{11}{c}{$1.0 \times 10^{-16}$} \\
	        Mean velocity of dust particles & MIV & km/s & \multicolumn{11}{c}{20} \\
	        Vapor phase production rate & MIV & - & \multicolumn{11}{c}{5} \\
	        Vapor temperature & MIV & K & \multicolumn{11}{c}{4000} \\
	        \midrule
	        Mean ion flux & SP & 1/cm$^2$s & \multicolumn{11}{c}{$1.08 \times 10^9$} \\
	        Yield efficiency & SP & - & \multicolumn{11}{c}{0.1} \\
	        Impact energy & SP & eV & \multicolumn{11}{c}{1000} \\
	        Porosity coefficient & SP & - & \multicolumn{11}{c}{0.35} \\
	        Microshadows coefficient & SP & - & \multicolumn{11}{c}{0.4} \\
	        \midrule
	        Photoionization lifetime & Exo & s & 600 & 2500 & 25000 & 6000 & 4000 & 8000 & 5000 & 20000 & 8000 & 20000 & 50 \\
	        Radiation acceleration & Exo & cm/s$^2$ & 5 & 5 & 5 & 15 & 25 & 5 & 5 & 5 & 5 & 5 & 5 \\
	        \bottomrule
	    \end{tabular}
	  		\label{tab:full_exo_base}
		\end{sidewaystable}

	\subsubsection{Photon flux and Photon-stimulated desorption (PSD)}

		PSD is initiated by the interaction of incoming photons with the surface, each photon possessing 
		the capability to eject atoms or molecules from a population of loosely bound volatile species. The 
		efficiency of this process is contingent upon the cross section for photon impact \cite{wurz1,killen3,wurz3}. 
		At perihelion, the incident photon flux is quantified as $3.1\times10^{16}$ cm$^{-2}$s$^{-1}$.

		The model quantifies the neutral particle flux resulting from PSD as:
		
		\begin{equation}
			\Phi_n = N_c \int \Phi_{\gamma}(E) \sigma(E) dE,
		\end{equation}

		where $\Phi_{\gamma}(E)$ denotes the energy-dependent differential photon flux, $\sigma(E)$ 
		the relative differential cross-section for desorption, $N$ the surface density of the regolith, and $c$ 
		the fraction of the specific neutral species being considered.

		The photon flux as a function of incidence angles is described by the following relation:

		\begin{equation}
			\Phi_n(\phi,\lambda)^\star = \Phi_n \cos(\phi) \cos(\lambda),
		\end{equation}

		with $\phi$ representing the longitude in local solar time and $\lambda$ the latitude, thereby factoring in 
		the geometric reduction of flux due to the angle of solar incidence.

		The energy distribution of the PSD process is modeled using a formula adapted from \citeA{johnson2002}:

		\begin{equation}
			f(E) = \beta(1 + \beta) \frac{EU^\beta}{(E + U)^{2+\beta}},
		\end{equation}
		
		in which $\beta$, the shape parameter, is set to 1 for our study, and $U$ denotes the threshold energy. 
		This value of $\beta$ is adapted from those presented in \citeA{johnson2002} to represent an energy cut-off 
		for this less energetic process. It is important to note that the PSD parameter subspace is not well constrained in the literature.
		In particular, the shape of the energy distribution may exhibit higher variance and interdependencies with other processes.
		Nevertheless, we consider our model representation sufficiently complex for the purpose of this study.
		Further exploration of these aspects is discussed in Section 6.

	\subsubsection{Micrometeorite fluxes and impact vaporization}

		MIV is a highly energetic surface-release process capable of releasing the entire surface material in a 
		given volume after a micrometeoroid impact (not only the volatile species). In the \citeA{mura1} model, 
		the MIV exosphere is simulated starting from a map of surface release particles at a given release energy, 
		represented as a Maxwellian distribution at 4000 K \cite{wurz1}.

		The vaporized material includes not only single elements but also molecules such as CaO, NaOH, NaO, 
		and others, resulting from the complex chemistry within the impact-produced cloud and a fraction of 
		condensed material that re-impacts the surface \cite{killen4, berezhnoy,Moroni2023}. However, for our initial 
		DNN analysis iteration, we assume these species have very short photolysis lifetimes, quickly breaking down into 
		their constituent elements without further energization. In other words, only single elements are released from 
		the surface.

		The distribution of this flux onto Mercury's surface is influenced by Mercury's velocity and its projection 
		onto the surface area where the flux is calculated, as illustrated in Figure \ref{fig:planet_orient}. We employ 
		a simple relationship between the angle of incidence and the modification of the mean flux onto the surface, 
		defined as:

		\begin{equation}
 			\Delta\phi_{MIV} = \frac{V_{mm}\cos{\beta_{Surf}}}{V_M},
		\end{equation}

		where $V_{mm}$ is the mean dust velocity, $V_M$ is Mercury's velocity and $\beta_{Surf}$ is the angle
		between Mercury's velocity vector and the surface normal vector. Consequently, the 
		incoming flux of dust particles varies between approximately $0.7\times10^{-16}$ and $1.4\times10^{-16}$ 
		gcm$^{-2}$s$^{-1}$ on the trailing and leading sides, respectively. At the chosen modal velocity of the 
		incoming flux, a constant vapor-phase production rate of about 5 is assumed in accordance with \citeA{cintala}:

		\begin{equation}
 			\frac{V_{x}}{V_{P}} = c + dv + ev^2,
		\end{equation}

		where $V_{x}$ and $V_{P}$ are the volumes of the released vapor and the impactors, respectively, $v$ is
		the velocity of the impactors, and $c$, $d$, and $e$ are constants. This simplification of the parameter subspace 
		for this complex vaporization process is deemed sufficient for our study.

		This approach yields outflows of surface matter ranging from $3.5\times10^{-16}$ to $7\times10^{-16}$ 
		gcm$^{-2}$s$^{-1}$. These values are estimated to be marginally smaller than those suggested 
		by \citeA{cintala} and two orders of magnitude smaller than those proposed by \citeA{pokorny}, 
		fitting within the parameter space of interest to modelers without overly emphasizing this omnipresent process.
		Additionally, this assumption is considered to present a more challenging scenario for the algorithm 
		as the surface composition is inherently reflected in the exospheric portion generated by MIV.

	\subsubsection{Ion precipitation and Ion sputtering (SP)}

		The SP is initiated by a flux of bombarding ions, predominantly comprising solar wind protons, which 
		efficiently ejects atoms/molecules from the surface regolith \cite{wurz1,mura0,wurz3, killen1}. The ion flux’s 
		impact is localized mainly in areas where the open magnetic field lines intersect the surface.

		In our model, the flux impacting Mercury’s surface is assumed proportional to the solar wind’s unperturbed 
		upstream flux of protons \cite{Massetti2003}, represented as:

		\begin{equation}
			\phi = C\rho_{sw}v_{sw},
		\end{equation}

		where $C$ denotes the ratio between the cusp area at the magnetic footprint and its corresponding area in 
		the undisturbed solar wind, set at 0.4 for our study. Here, $\rho_{sw}$ is the solar wind density (60 cm$^{-3}$), 
		and $v_{sw}$ is the solar wind velocity (450 km s$^{-1}$). The calculated flux impacting the surface is 
		$1.08\textrm{x}10^{9}$ cm$^{-2}$s$^{-1}$.

		To derive the flux for individual species, we employ the equation from \citeA{mura1}:

		\begin{equation}
			\frac{d\Phi_n}{dE_e} = Yc \int_{E_{\text{min}}}^{E_{\text{max}}} \frac{d\Phi_I}{dE_i} f_S(E_e, E_i) dE_i,
		\end{equation}

		where $Y$ is the yield of the process, $c$ the surface relative abundance of the species, $\Phi_I$ the ion flux, 
		$\Phi_n$ the neutral flux emitted from the surface, $E_i$ the impact energy, $E_e$ the energy of the ejected 
		particles, and $f_S$ an empirical model for the energy distribution of ejected particles, defined as:

		\begin{equation}
			f_S(E_e, T_m) = c_n \frac{E_e}{(E_e + E_b)^3} \times \left[ 1 - \left( \frac{E_e + E_b}{T_m} \right)^{1/2} \right],
		\end{equation}

		with $T_m$ as the maximum transmitted energy, $c_n$ the normalization constant, and $E_b$ the surface binding 
		energy of the ejected species. $T_m$ is:

		\begin{equation}
			T_m = E_i \frac{4 m_1 m_2}{\left( m_1 + m_2 \right)^2} ,
		\end{equation}

		where $E_i$ is the impact energy, taken as constant 1000 eV, $m_1$ and $m_2$ are the masses of the impinging
		ion and the extracted particle, respectively.

		For this investigation, we assume the same yield efficiency for all species, defined as:

		\begin{equation}
			Y = Y_0 \times \gamma_{por} \times \gamma_{ms} ,
		\end{equation}

		where $Y_0 = 0.1$ is the base yield, and the coefficients $\gamma_{por} = 0.35$ and $\gamma_{ms} = 0.4$
		represent the effects of the regolith's porosity and the microshadows within it, respectively \cite{Jaggi2024}. 
		This is a rough approximation since each element has a different binding energy, resulting in different yields. 
		Nevertheless, by considering this low yield, we aim to account for its overall reduction due to the aformentioned 
		microscopic effects. This overall low sputtering effect is a deliberate choice to complicate the prediction of surface composition 
		by DNN algorithms in high-latitude regions receiving solar wind precipitation. 
		The angular distribution around the normal direction of the surface is taken as $\cos^2{(\alpha_n)}$.

	\subsubsection{Dynamics of the Exosphere}

		The dynamics of the exosphere, as simulated in the \citeA{mura1,mura2} model, encompass the movement and 
		behavior of particles after they have been released into the exosphere, including their interactions,
		trajectories, and eventual fate. Factors such as gravitational influences, electromagnetic forces,
		and collisions are examined to understand how they shape the structure and composition of the
		exosphere.

		Once in the exosphere, each elemental species follows ballistic trajectories under the influence of 
		Mercury’s gravity. These particles are also subject to conditions that define their mean lifetime in the exosphere 
		before impacting the surface or undergoing photoionization. When an exospheric particle is ionized, it 
		is no longer simulated, making the ionization process a net loss to the exosphere. 

		Another relevant effect included in the model is solar radiation pressure, which tends to push neutral elements 
		away from the direction of incoming sunlight, effectively propelling them toward the night side of the planet. 
		This movement is not uniform across all species; it varies depending on the physical properties of the particles, 
		such as their mass and effective cross-section, which influence how much momentum they absorb or reflect 
		from solar photons. The parameters for each of the elemental species used in our exospheric model 
		are listed in Table \ref{tab:full_exo_base}.

\subsection{Generation of the Datasets}

		We conduct separate simulation runs for each of the four primary surface release processes, reproduced 
		for each distinct species. Following these individual simulations, we aggregate the outcomes to compose the 
		overall exosphere. This process essentially involves summing the resulting individual exospheres 
		generated for each species, without considering interactions between the various processes. For instance, 
		we do not account for potential competition among processes for a finite pool of particles at the surface. 
		Similarly, the exospheres for different species are treated as non-interacting entities. 

		The number of particles launched in the Monte Carlo simulation significantly affects the accuracy of the results. 
		A higher number of particles better captures the statistical behavior of the populations. However, the need for 
		more computational resources and time increases with the number of particles. For our purposes, considering 
		the coarse resolution and the high number of simulations required, we launch 50,000 particles for each simulation 
		run.

		The exospheric grid is a virtual representation of space around the planet, divided into discrete cells, serving 
		as the framework for tracking particle positions and movements. For our study, we have extended the surface 
		2D grid in 100 km altitude steps around the planet to a final altitude of 5000 km.

		This section further describes the comprehensive process of dataset generation, detailing how we simulate 
		measurements and observations that mimic real-world exospheric data.

		\subsubsection{Measurements and Observations Creation}

		As explained above, the creation of our datasets starts from a randomly generated surface elemental 
		composition based on realistic mineralogies, establishing the ground truth for each dataset. We then simulate 	
		environmental effects to craft a static representation of the exosphere at a specific moment, taking into 
		account the necessary physical and chemical processes. 

		Next, we strategically define various positions within the exosphere to place a virtual sensor, aiming to comprehensively 
		provide measurements of its constituents. This conceptual sensor operates without considering the complexities 
		and potential inaccuracies introduced by real-world sensor characteristics such as detector noise and sensitivity 
		limitations. To create a comprehensive and static snapshot of the exosphere, measurements across the simulated 
		exosphere are conducted simultaneously.

		The positions in the exosphere are selected to maximize data diversity and relevance, considering factors like 
		altitude, latitude, and environmental conditions. To balance the required spatial variability and facilitate 
		the reconstruction of complete surface maps from DNN predictions, we align our virtual sensor positioning in the 
		exosphere directly above each surface grid tile's center. These measurements collectively form what we refer to 
		as an observation or a data subset. Each observation/subset comprises 648 measurement data points per altitude
		level. Every data point within an observation is a vector that encapsulates the measured densities of all neutral 
		species present in the exosphere as per our simulation's setup. To each data point there is also the corresponding 
		ground truth vector formed from the elemental fractions of the surface tile below.

		In terms of altitude coverage, we define two types of data subsets. The first is a constant altitude observation, with 
		different subsets at altitudes ranging from 200 km to 1500 km. For the second type, the augmented subset, we 
		applied data augmentation techniques to enhance our datasets, especially for training. This involved incorporating 
		measurements from a range of altitudes, including 200, 500, 800, 1100, 1400, 1700, and 2000 km for a total of
		4,536 data points per subset. This approach not only increases the dataset volume but also introduces a variety 
		of exospheric conditions, allowing the algorithm to develop a nuanced understanding of how measurements at 
		different altitudes correlate with surface compositions.

	\subsubsection{Training, Validation and Test Datasets}

		We produce three distinct types of datasets: training, hold-out validation, and testing datasets. This subdivision 
		is a fundamental practice in machine learning, ensuring that the algorithm is trained on a diverse set of data, 
		validated for accuracy, and finally tested for generalization to unseen data. Different subsets of generated 
		exospheres are used for training, validating, and testing the DNN. For the training datasets, multiple 
		observations/subsets representing different exospheres are aggregated, enhancing the dataset’s size, complexity, 
		and variability. Conversely, the hold-out validation and testing datasets are each derived from individual 
		observations/subsets to maintain representability of particular planetary surface-exosphere pairs. 

		Given the mission-specific orbital characteristics of the BepiColombo/Mercury Planetary Orbiter and the 
		enhanced measurement capabilities of the STROFIO mass spectrometer within the SERENA instrument suite at lower
		altitudes, we define a baseline constant altitude subset at 500 km. This altitude, corresponding to the periherm 
		(closest approach) of the MPO, is chosen for its potential to yield accurate measurements with an improved 
		signal-to-noise ratio. In our training campaign we examine both the baseline dataset type with examples only 
		at altitudes of 500 km (648 examples per subset) and the augmented dataset type (4,536 examples per subset), 
		the latter of which provides the MLP DNN with the capability to make predictions at diverse altitudes. 

		The training phase demands a complex strategy that captures a larger part of the data distribution to 
		train the algorithm to approximate the relationships between the employed physical processes accurately. 
		To address this, we aggregate multiple data subsets or observations to form the training dataset, with 
		the number of included subsets ranging from 10 baseline observations (yielding a total of 6,480 data points) to 
		300 augmented observations (resulting in a staggering 1,360,800 data points). The multiple training sets used in 
		our study vary in size and complexity and present different empirical distributions to the DNN. Training the 
		DNN on each set produces different estimators of the processes, each tailored to the particular empirical distribution 
		of its training set. The construction of a representative training dataset plays a crucial role in ensuring the 
		resulting estimator of physical processes closely approximates the actual data distribution. Additionally, increasing 
		the training set size reduces the probability of the algorithm becoming biased towards a non-representative 
		smaller dataset distribution.

		For the validation and testing phases of our DNN algorithms, we adopt a different strategy. Hold-out validation 
		involves using a single validation dataset to monitor the trained network during each training epoch. This process 
		helps identify when the algorithm begins to overfit the training data, thereby reducing its generalization capabilities. 
		The validation set is also crucial during hyperparameter tuning to evaluate the algorithm’s ability to generalize 
		effectively. At the same time, the test datasets consist of examples (data points) that are never seen during training, 
		ensuring that the MLP DNN algorithm’s performance evaluation is not biased by improvements in accuracy due to 
		evaluation on previously seen examples.

		For validation and test dataset purposes, employing single data subsets allows for straightforward prediction and
		reconstruction of surface maps corresponding to individual surface-exosphere simulations. Our hold-out validation 
		dataset is selected at the MPO's periherm altitude of 500 km (baseline subset) to maintain consistency in evaluation 
		conditions. 

		The test datasets are designed to thoroughly assess the algorithm's predictive capabilities under varied conditions. 
		To ensure a thorough assessment of the algorithm's performance, we generate a total of 135 test datasets derived from 
		15 distinct exosphere simulations, each at a constant altitude ranging from 200 km to 1500 km. This diversified
		testing ground allows us to explore the algorithms' responsiveness to altitude variations. Such an approach facilitates a 
		comprehensive analysis and evaluation across a broader spectrum of exospheric altitudes and conditions, enabling 
		us to more accurately gauge the algorithm's efficacy and robustness in predicting the surface elemental composition. 
		A more detailed description of the datasets and their underlying surfaces is given in Appendix B.

	\subsubsection{Feature Selection and Engineering}

		The final step is to supplement the datasets with additional metadata, or features, from these observations, 
		such as geometrical information, to serve as auxiliary inputs for the algorithms. The effectiveness of deep neural 
		network algorithms in modeling complex relationships within data is significantly influenced by the selection and 
		engineering of input features. Carefully chosen or engineered, these features, describing each data point within all 
		three types of datasets, can enhance the algorithm's ability to discern patterns and relationships, thereby improving 
		its overall performance.

		In our study, we have considered and incorporated a range of additional features to enrich our datasets:

		\begin{itemize}
			\item \textbf{Altitude of Measurement}: This feature is critical for capturing altitude-specific dynamics, enabling 
			the algorithm to identify how the distribution of neutral species changes with altitude relative to their source points 
			on the surface. Both the actual altitude and a logarithm of the altitude were tested as features. The latter 
			engineered feature is intended to highlight non-linear altitude effects on the measured parameters, providing 
			another layer of depth to the altitude-related analysis.

			\item \textbf{Logarithm of Exospheric Density}: By applying the base 10 logarithm to the exospheric density of 
			each species, we introduce a constraint that aids the algorithm in exploring nonlinear relationships, acknowledging 
			the exponential decrease in density with altitude.

			\item \textbf{Subsolar Angle}: Represented as either the value of the angle $\phi$ directly, or as $\cos(\phi - 180)$, 
			this feature helps differentiate between exospheric populations on the dayside, nightside, and the transitional 
			terminator regions, enhancing the model's spatial awareness.

			\item \textbf{Latitude Dependency}: Using $\sin(\gamma)$, this feature allows the algorithm to account for 
			latitude-specific phenomena, such as ion sputtering, which vary across different latitudinal zones.

			\item \textbf{Proton Flux Virtual Data}: Integrating virtual proton precipitation measurements, this feature hints at 
			sputtering effects on the surface, offering a proxy for understanding underlying ion induced physical processes.
		\end{itemize}

		Each data point in our datasets is defined by combinations of these features, forming distinct feature sets that 
		illuminate to the algorithm various aspects of the exosphere's behavior. The compilation of these feature sets 
		is crucial for unraveling the capabilities of neural networks in predicting surface compositions and contributing 
		insights into the mechanisms governing particle release into the exosphere.

\section{Results}

In this section, we show the findings of our investigation, which are divided into two distinct phases to provide a 
comprehensive understanding of our study's outcomes. The first, training phase focuses on the configuration and optimization 
of the deep neural network. This entails a systematic exploration of the hyperparameter space and other method 
characteristics to identify the optimal settings that enhance the DNN's ability to model the data accurately. The second, 
testing phase evaluates the performance of the DNN, now finely tuned with the optimal hyperparameter configuration, 
in interpreting and making predictions on data derived from unseen during learning examples.

\subsection{Training Phase and DNN Finalization}

	An extensive training campaign was undertaken to explore both the empirical distribution represented 
	in the training datasets, and the hyperparameter space of the neural network architecture. This effort 
	aimed to develop an accurate estimator that demonstrates optimal generalization capabilities by closely
	approaching the true data generating distribution. In this section we outline the most important findings of 
	this investigation phase to ultimately refine and finalize the components of the multilayer perceptron deep 
	neural network. The complete training campaign is detailed in Appendix C.

	\subparagraph{Eliminating Skewed Predictions}\hfill

		Initial analysis showed that accuracy metrics for predicting surface elemental composition were 
		skewed by the high prevalence of oxygen (O$_2$) and silicon (Si). To address this, we excluded these 
		elements from the prediction vector, adjusting the model to focus on the normalized proportions of the 
		remaining nine elements. This adjustment improved the model's relevance and performance by aligning 
		with our study's objectives more effectively.

	\subparagraph{Training Set Size and Data Augmentation}\hfill

		We analyzed MLP DNN performance in relation to the expansion of the training dataset size with training 
		sets ranging from 10 to 200 unaugmented baseline data subsets, observing that larger datasets improved 
		generalization accuracy. Additionally, in order to better approximate the true data-generating distribution, 
		we augmented our training datasets with examples varying in altitude, enhancing representability. This strategic 
		choice expanded our dataset to 300 augmented subsets, totaling 1,360,800 examples, significantly improving 
		the model's robustness and predictive accuracy across altitude-specific inputs.

	\subparagraph{Learning Curve Examination for Optimal Training Duration}\hfill

\begin{figure}[h!]
			\centering
			\includegraphics[scale=0.7]{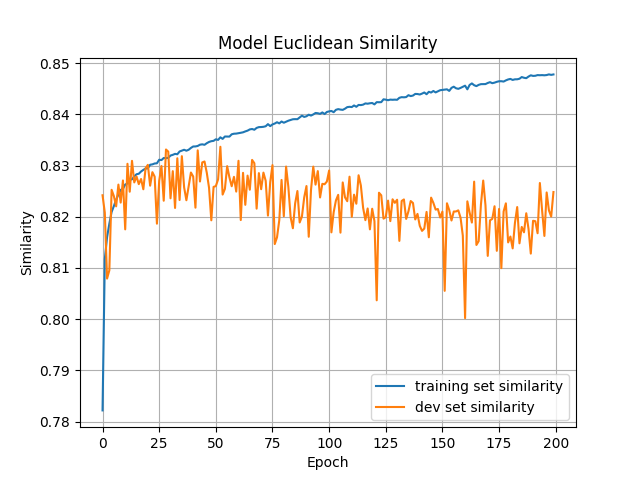}
			\caption{Learning curve for the MLP DNN training. The blue and orange curves show the evolution 
			of the average prediction similarity of the full training dataset 
			(300 subsets, 1,360,800 data points) and the hold-out validation dataset (1 subset, 
			648 data points) respectively.
			}
  			\label{fig:learn_curve_300}
		\end{figure}

		Our examination of the MLP DNN's learning curves aimed to identify the optimal training duration to avoid 
		overfitting, in line with the guidance provided by \citeA{bengio}. Analysis indicated that predictive performance 
		on the validation dataset peaked at 40 epochs, as shown in Figure \ref{fig:learn_curve_300}. While longer training 
		durations, up to 200 epochs, continued to align the model to the training dataset, the best balance between 
		training and inference accuracy was achieved at 40 epochs, suggesting this as the optimal training duration.

	\subparagraph{Selected Feature Set}\hfill

		After rigorous testing and evaluation, the feature set that emerged as superior, offering the most 
		consistent and highest accuracy, comprised of the following features: 

		\begin{itemize}
			\item Logarithmic transformations of elemental species exospheric densities.

			\item Logarithm of the altitude at which measurements were taken.

			\item Sun incidence angle.

			\item Presence of H+ ions arriving through open field lines.

			\item Local time.

			\item Latitude.
		\end{itemize}

	\subparagraph{Hyperparameter Optimization and DNN Structural Components Finalization}\hfill

\begin{figure}[h!]
			\centering
			\includegraphics[width=\linewidth]{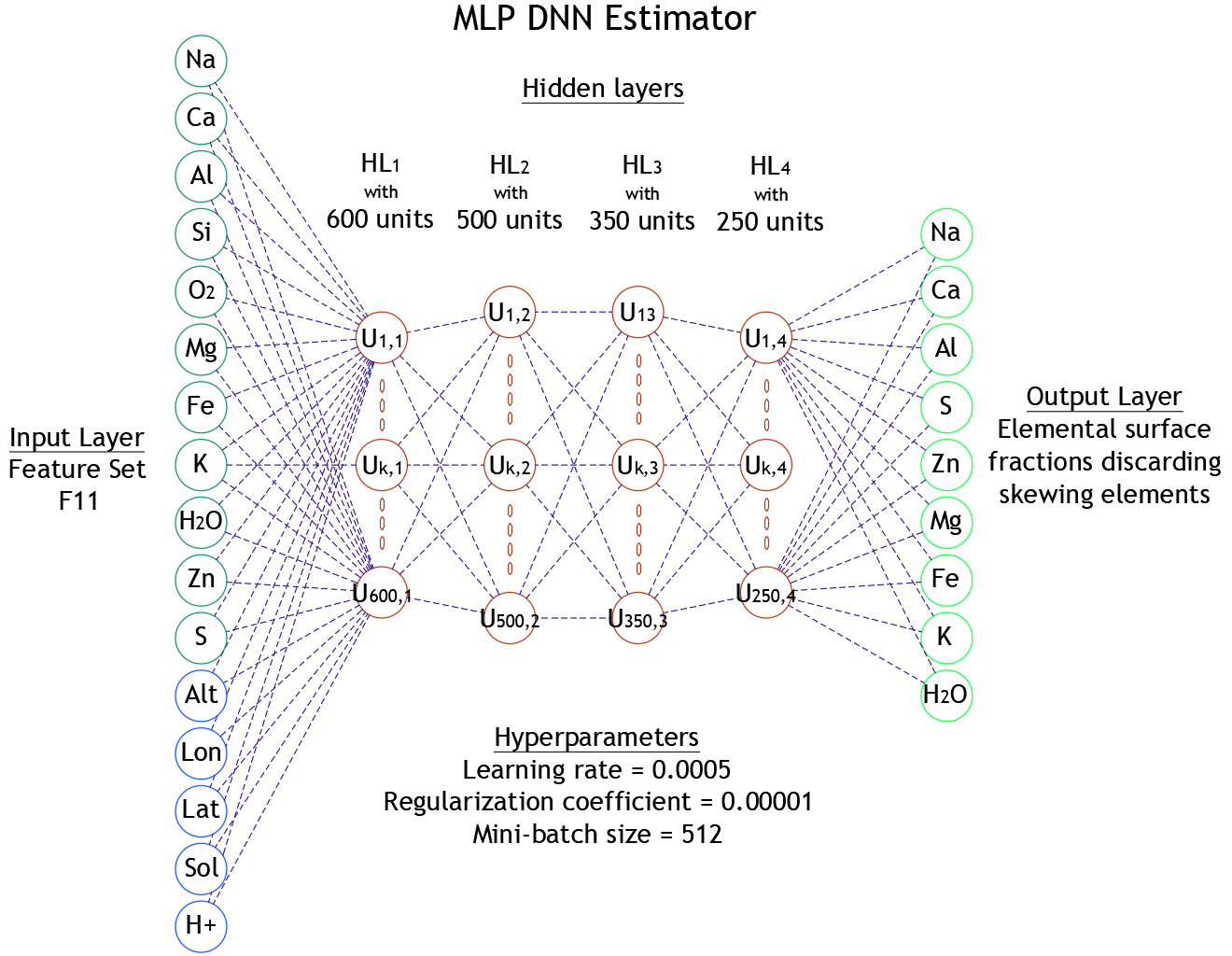}
			\caption{Finalized multilayer perceptron deep neural network. The input layer consists of the features
			collected in feature set F11, the output layer is adjusted to remove the skewing predictions towards
			better estimating the O$_2$ and Si species. There are four hidden layers with 600, 500, 350, and 250
			neurons respectively.
			}
  			\label{fig:mlp_final}
		\end{figure}

		An extensive hyperparameter optimization effort resulted in the selection of the final MLP DNN architecture, 
		consisting of a four-layer structure with 600, 500, 350, and 250 neurons in each layer respectively 
		(Figure \ref{fig:mlp_final}). The regularization coefficient was optimized to a higher value of $1.0\times10^{-5}$ 
		to improve generalization, while the learning rate was finely tuned to $0.5\times10^{-4}$. Training was 
		conducted in mini-batches of 512 examples, identified as near-optimal through our optimization process. 

		The combined efforts of hyperparameter exploration, architectural fine-tuning, and dataset augmentation 
		have significantly propelled our model's performance and its ability to generalize from the empirical distribution 
		to the true underlying data-generating distribution—by as much as a total of 10\% in ES4 and up to 30\% in 
		R-squared. This optimized architecture, along with structural parameters and algorithmic characteristics 
		refined during our comprehensive training campaign, ensures that the MLP DNN is a robust model for our 
		sophisticated predictive tasks.

\subsection{Testing Phase}

	The ultimate evaluation of our multilayer perceptron deep neural network algorithm's performance 
	hinges on its ability to accurately predict surface compositions and reconstruct elemental surface 
	maps using datasets it has not previously encountered. Our research incorporates two distinct test campaigns 
	designed to assess the MLP network's predictive prowess. These campaigns were structured to apply the final 
	network, fine-tuned with an extensive training set comprising 300 augmented subsets, across test datasets 
	derived from a variety of altitudes not previously seen during training.

	The scope of these test campaigns is broad, focusing not only on aggregate performance metrics 
	across the entire dataset but also on detailed analyses for individual elemental species. This includes 
	a thorough examination of residuals to identify any systematic errors or biases in predictions, with the final goal
	of reconstructing the surface composition maps for each species. This process entails a visual comparison between 
	the original, or ground truth, maps and the ones predicted by our algorithm.

	\subsubsection{Preliminary Test Campaign}

		In our preliminary test campaign, we embarked on a performance evaluation using single-simulation 
		test datasets derived from 15 unique surface compositions, each leading to distinct exospheres. This 
		approach encompassed data from both the dayside and nightside, allowing for a robust  examination 
		of the MLP DNN algorithm’s predictive accuracy and its capability in reconstructing surface elemental 
		maps under varying conditions.

		The campaign tested the algorithm's performance across a spectrum of altitudes ranging from 200 km 
		to 1500 km. This setup provided a rich dataset for analysis, comprising 15 sets of predictions for each 
		of the 9 altitude levels, culminating in a total of 135 complete prediction sets. These predictions detailed 
		the fractional composition of nine elements across the surface grid tiles, facilitating the reconstruction 
		of elemental maps for the 15 different surfaces from measurements at each altitude level.

\begin{figure}[h!]
			\centering
			\includegraphics[width=\linewidth]{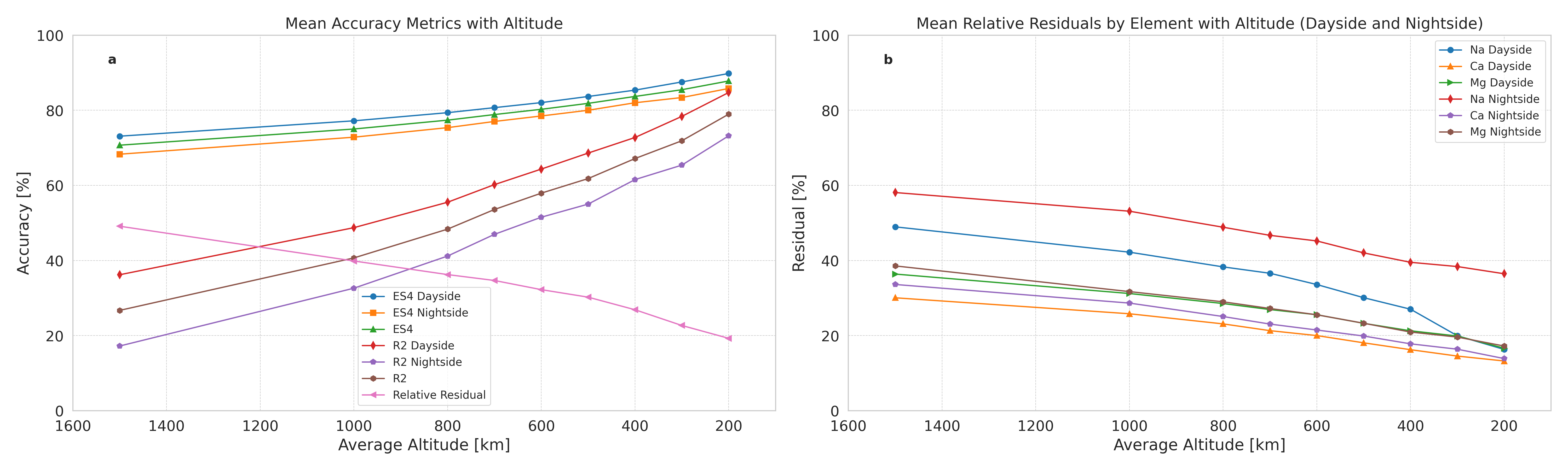}
			\caption{Preliminary test campaign. Panel \textbf{a} shows the mean accuracy (dayside, nightside, and overall) 
			and relative residuals metrics of the MLP DNN predictions on the 15 test surfaces. In panel \textbf{b} are 
			displayed the dayside and nightside mean relative residuals for the elements Na, Ca, and Mg.
			}
  			\label{fig:predavg_comb}
		\end{figure}

		We utilized our suite of performance metrics, including the average ES4, R-squared, absolute, and relative 
		residuals, to evaluate the predictions and reconstructions systematically. These metrics were plotted 
		against the measurement altitudes to analyze the model's performance comprehensively, depicting them 
		for the overall predicted output, individual elemental species, and separate analyses for dayside and 
		nightside predictions (Figure \ref{fig:predavg_comb}).

		The results of this preliminary testing suggested that predictions were generally more precise for the dayside, 
		likely a consequence of particle movements influenced by solar radiation pressure. This was especially true
		for volatile species, such as sodium, on which we observed a notable discrepancy in predictive accuracy
		with respect to refractory ones, like magnesium and calcium.

	\subsubsection{Main Test Campaign}

		We note that Mercury has a unique rotational and orbital dynamics, particularly its 3:2 orbit-spin 
		resonance. This results in the nightside hemisphere becoming the dayside in every following orbit. Thus, by 
		using a combination of data from two consecutive years, we can fill the gap in the predictions of the volatile 
		species, as leveraged in the main test campaign. This approach utilized double-simulation 
		compound predictions, focusing on the same 15 surface compositions from the preliminary campaign 
		but observed at two consecutive perihelia. During these two periods, different halves of Mercury's surface 
		were illuminated by the Sun, allowing for comprehensive daylight observation of the entire planet over 
		the two simulations. For this campaign, predictions specifically targeted sunlit surface tiles, enabling an 
		in-depth analysis of surface compositions that were previously on the nightside in the initial test phase.
		Measurements for this campaign were again taken at a range of altitudes from 200 to 1500 km.

		A significant outcome of the combined odd-even orbit campaign was the improved accuracy in predicting 
		volatile species' fractions, aligning more closely with the refractory species' predictions observed in 
		the preliminary campaign. This enhancement in predictive accuracy for volatiles under daylight conditions 
		underscores the importance of solar illumination in accurately assessing surface compositions.

\begin{figure}[h!]
			\centering
			\includegraphics[width=\linewidth]{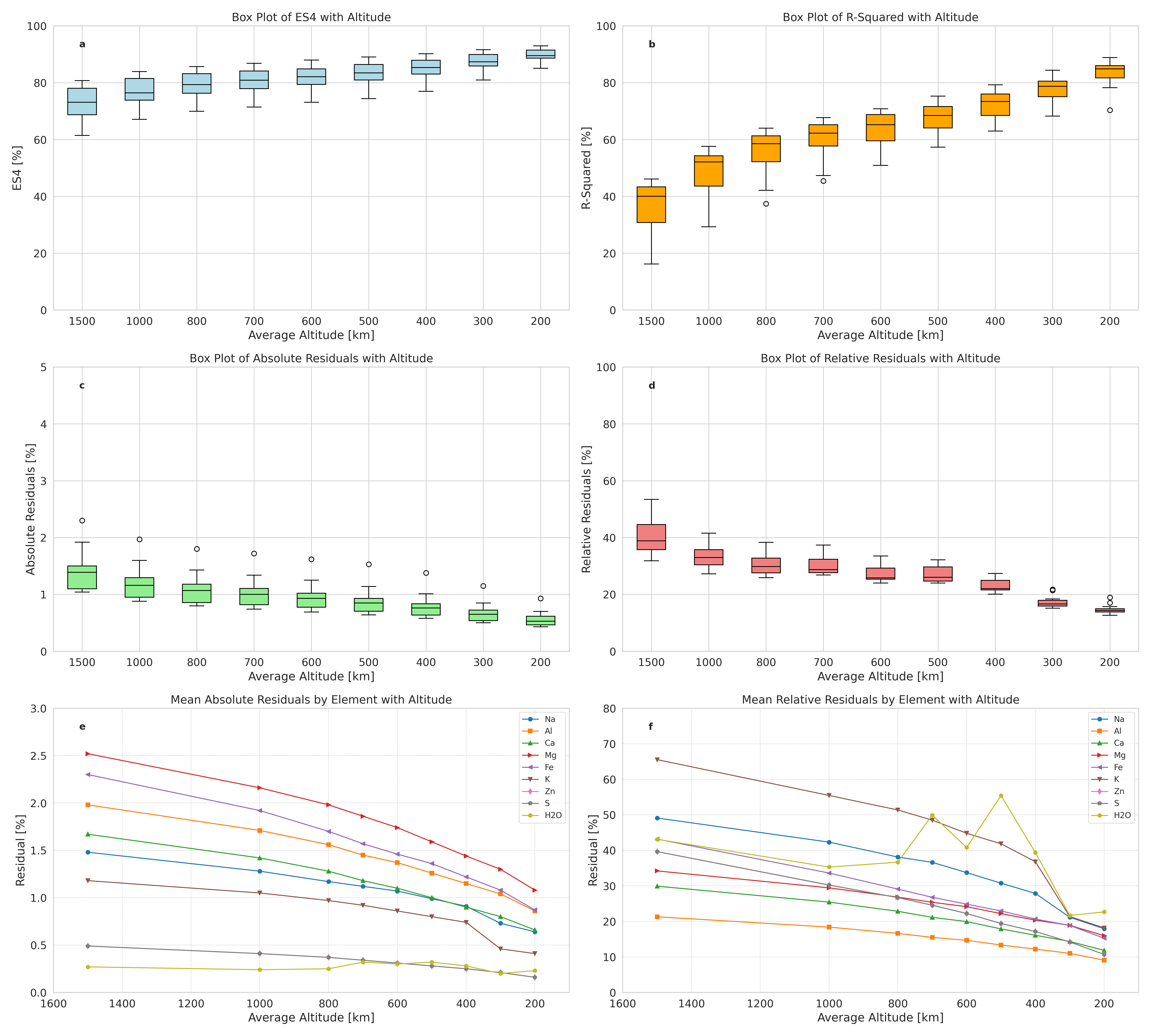}
			\caption{Box plots of the accuracy metrics (panels \textbf{a} and \textbf{b}) and the mean residuals 
			(panels \textbf{c} and \textbf{d}) of the MLP DNN predictions on the 15 test surfaces of the main test 
			campaign. Panels \textbf{e} and \textbf{f} give respectively the absolute and relative residuals for all 
			predicted elements averaged over the 15 predicted surfaces.
			}
  			\label{fig:predavg_comb_6x}
		\end{figure}

		The main campaign demonstrated a notable increase in overall prediction and map reconstruction 
		accuracy, with the average ES4 metric reaching approximately 89.70\% and the average R-squared metric reaching 
		83.41\% at the lowest altitude of 200 km (Figure \ref{fig:predavg_comb_6x}). This accuracy diminished at 
		higher altitudes, attributed to the exosphere's dynamic nature and the increased complexity in tracing back 
		exospheric particles to their originating surface tiles. There was, however, a marked improvement compared to
		the preliminary findings, which highlights the efficacy of considering Mercury's solar exposure in enhancing 
		predictive models' accuracy. By focusing solely on the dayside observations across two perihelia, the campaign 
		effectively capitalized on optimized conditions for surface composition reconstruction. The box statistical plots shown 
		on Figure \ref{fig:predavg_comb_6x} suggest also reduction in the range of prediction accuracies and residuals 
		with decrease in altitude.

		Going deeper in the detailed statistics of the predictions by the MLP DNN for each elemental species, we can
		observe its tendencies in the box plots of the absolute and relative predictions for Aluminium, Calcium, and 
		Sodium (Figure \ref{fig:boxres_alcana}). Box plots for the remaining elements are provided in Appendix D. 
		Almost all elements have good prediction statistics at the lowest altitude, with the median relative residual of 
		Aluminium particularly impressive at only 8.69\% at 200 km. The algorithm has more difficulties with Sodium, 
		with its median relative residual at 17.86\% at 200 km, which is nevertheless a good result. The robustness in 
		the prediction of the refractive elements is present throughout the altitudes, even up to 1500 km, where the 
		median relative residuals of Aluminium is 21.18\%, while that for Calcium is 30.76\%. While the range of the 
		prediction errors for the different types of surface elements is impressive throught the altitudes.

		\begin{figure}[h!]
			\centering
			\includegraphics[width=\linewidth]{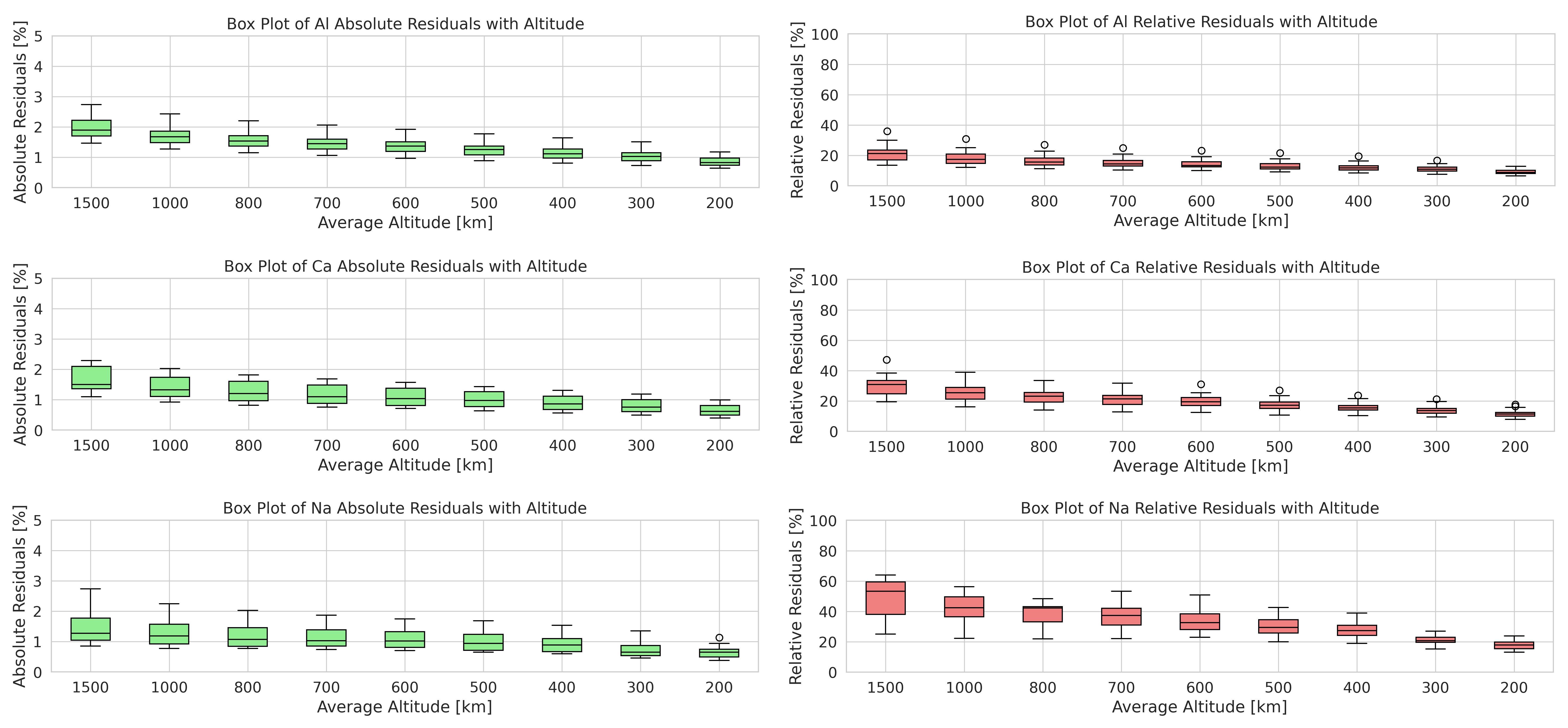}
			\caption{Box plots of the residuals metrics of the MLP DNN predictions on the 15 test surfaces of the main test 
			campaign for the elements Aluminium, Calcium, and Sodium. 
			}
		  			\label{fig:boxres_alcana}
		\end{figure}

\begin{figure}[h!]
			\includegraphics[width=\linewidth]{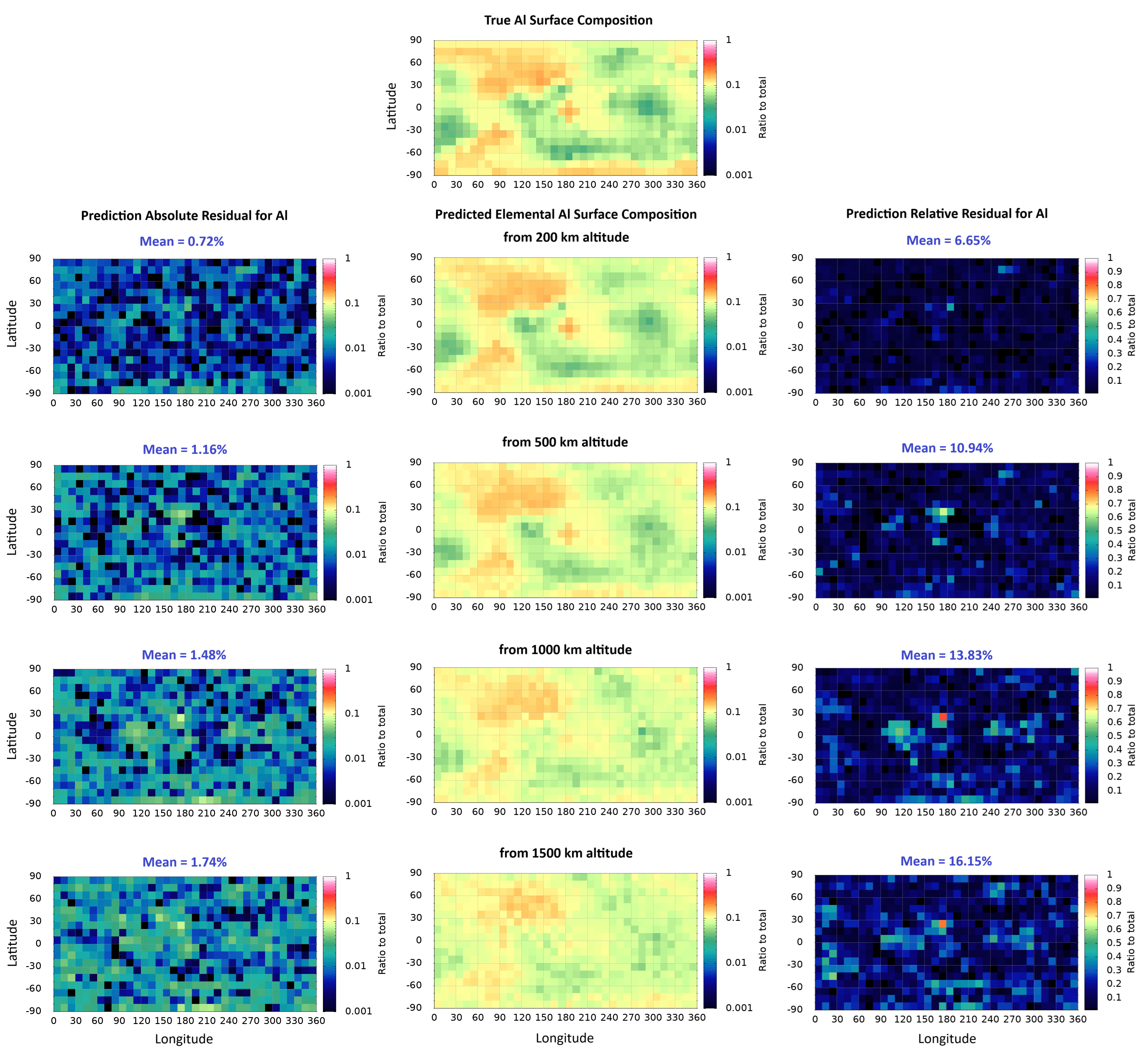}
			\caption{
			Main test campaign - MLP DNN map reconstructions of the same sample Aluminium 
			surface composition (test set number 2). Dayside only inputs of two simulated exospheres
			from consecutive Mercury perihelia at different altitude levels (200, 500, 1000, and 1500 km).  The top-most
			map shows the "ground truth" surface composition. The maps in the middle below it are the predicted
			fractions for this element. The panels on the left show the absolute residuals to the "ground truth", while
			on the right are the relative residuals.
			}
		  	\label{fig:c_map_recon_al}
		\end{figure}

\begin{figure}[h!]
			\includegraphics[width=\linewidth]{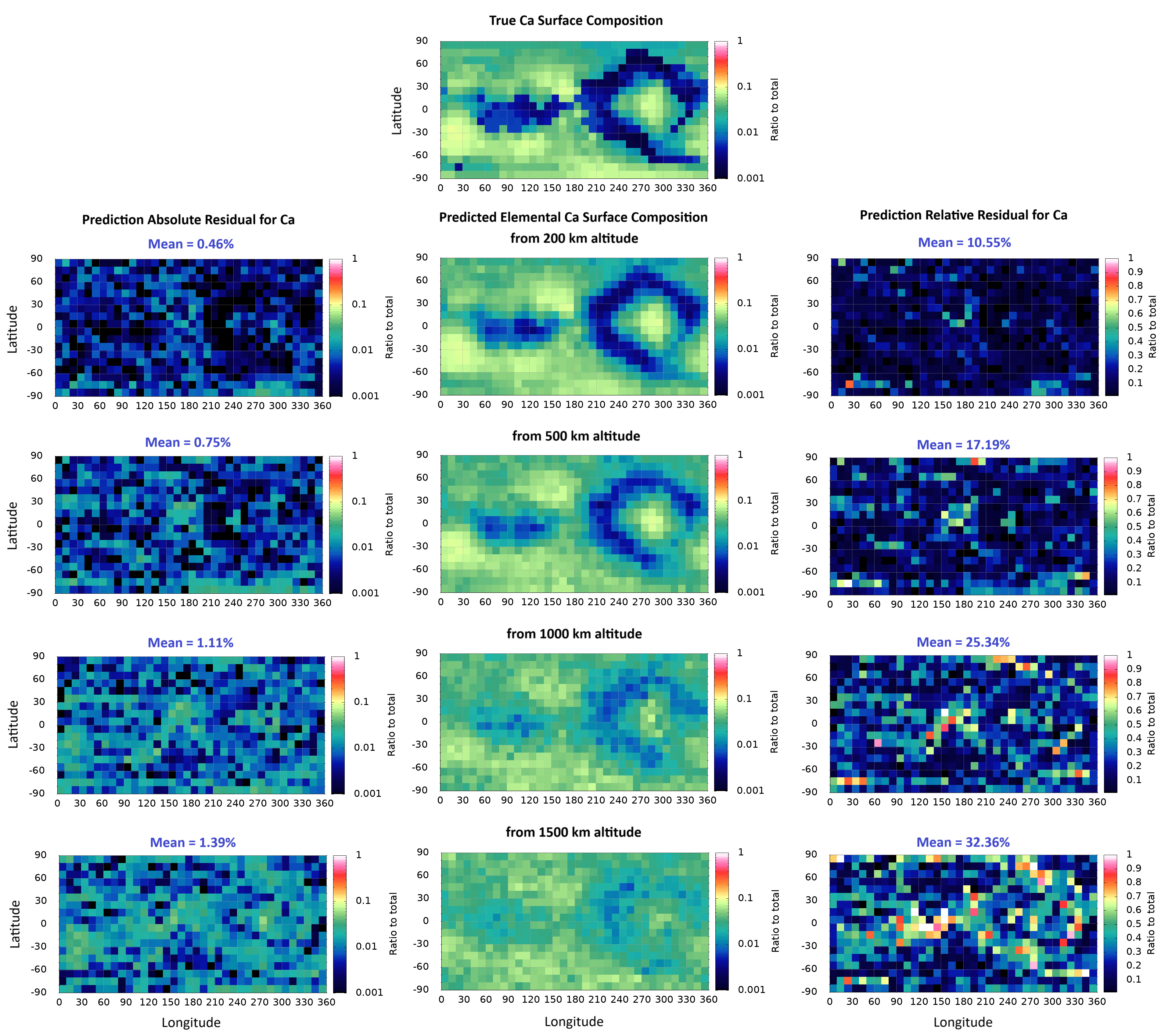}
			\caption{
			Main test campaign - MLP DNN map reconstructions of the same sample Calcium 
			surface composition (test set number 2). Inputs are coming from the dayside of two simulated exospheres
			from consecutive Mercury perihelia at different altitude levels (200, 500, 1000, and 1500 km).  The top-most
			map shows the "ground truth" surface composition. The maps in the middle below it are the predicted
			fractions for this element. The panels on the left show the absolute residuals to the "ground truth", while
			on the right are the relative residuals.
			}
		  	\label{fig:c_map_recon_ca}
		\end{figure}

\begin{figure}[h!]
			\includegraphics[width=\linewidth]{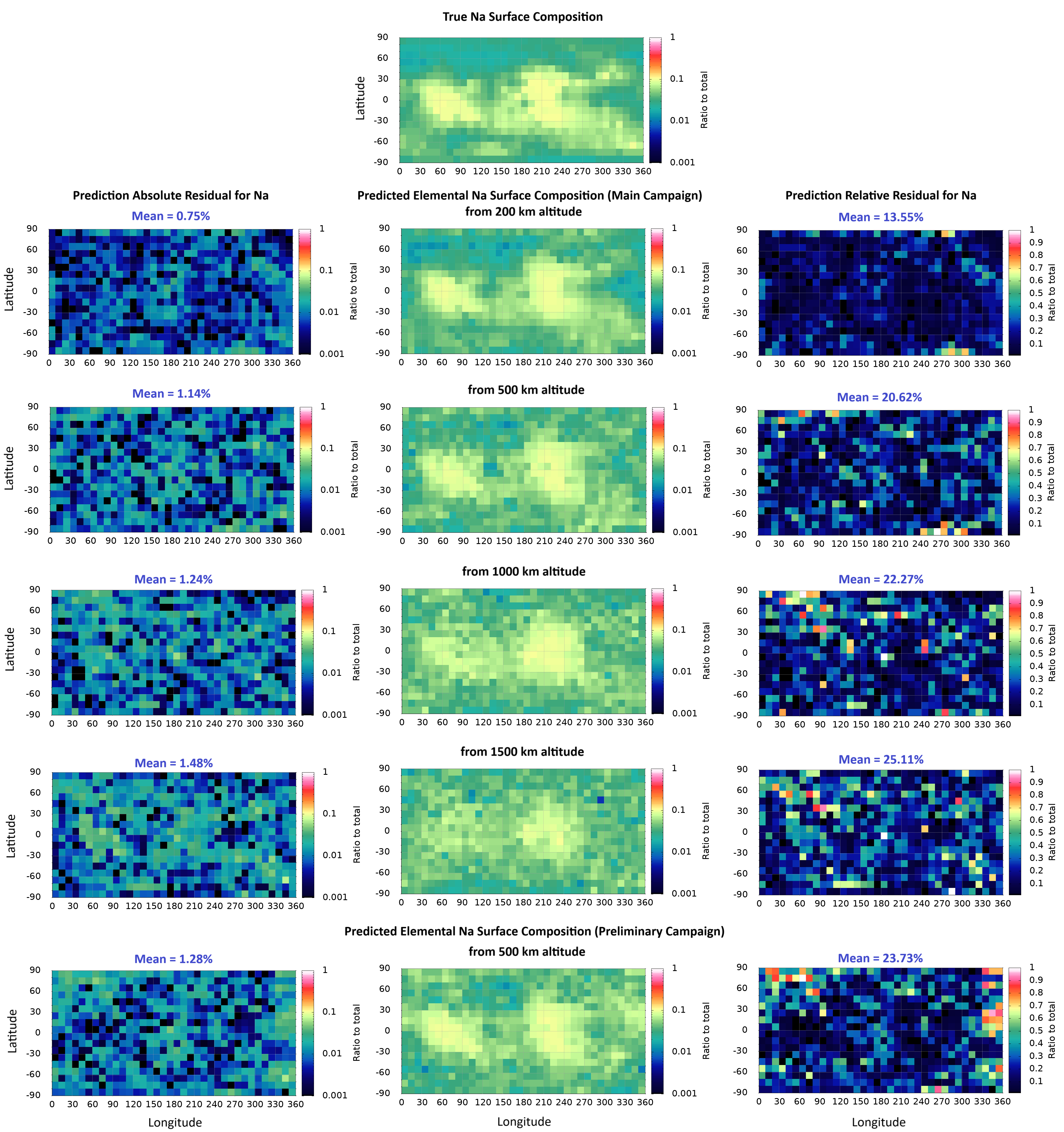}
			\caption{
			Main test campaign - MLP DNN map reconstructions of the same sample Sodium 
			surface composition (test set number 2). Dayside only inputs of two simulated exospheres
			from consecutive Mercury perihelia at different altitude levels (200, 500, 1000, and 1500 km).  The top-most
			map shows the "ground truth" surface composition. The maps in the middle below it are the predicted
			fractions for this element. The panels on the left show the absolute residuals to the "ground truth", while
			on the right are the relative residuals.
			Eclipsed areas in the preliminary campaign (bottom-most panels, 0-90 and 270-360 Lon) are markedly 
			more accurately reconstructed in these dayside only combined maps of the main campaign.
			}
		  	\label{fig:c_map_recon_na}
		\end{figure}

		Additionally, our visual comparisons from the map reconstructions (Figures \ref{fig:c_map_recon_al} to 
		\ref{fig:c_map_recon_na}) highlighted the algorithm’s strengths and weaknesses in predicting different 
		elemental distributions. 

		The discrepancy in predictive accuracy between refractory species, such as Aluminium (Figure \ref{fig:c_map_recon_al}) 
		and Calcium (Figure \ref{fig:c_map_recon_ca}), and volatile ones, like Sodium (Figure \ref{fig:c_map_recon_na}), 
		is still present in the predictions of the particular test set 2 shown on the figures. However, there is a noticable 	
		improvement by the daylight only predictions of the main testing campaign, compared to the one from the 
		preliminary campaign, which included the night side prediction. This is shown on the map reconstructions of Sodium 
		(test set 2), where the bottom-most panels in Figure \ref{fig:c_map_recon_na} show that increased errors on the 
		night side (longitudes 0-90 and 270-360) and an average relative residual of 23.73\% from the preliminary test, 
		directly compared to the average of the main campaign at 500 km (third row from top) where the average 
		relative residual is reduced to 20.62\%.

		From a closer examination of the reconstructed maps, we can clearly see that certain large scale patterns in
		the surface composition of all species are recognized from as far as 1500 km, even if details are missed by the 
		MLP at that large distance. The algorithm starts to resolve with a relatively good accuracy at altitudes of 500-800 km. 
		This is especially true for the refractory species (Ca, Al, etc.), and going closer to the planet (down to 200 km) 
		produces the most accurate predictions and reconstructions, even for volatiles (Na). The full set of reconstructed
		maps for this particular test set number 2 in our main test campaign is provided in Appendix D.

\section{Discussion}

\subsection{Methodological Framework and Proof-of-Concept Aims}

	This research presents a novel approach for deducing Mercury's surface composition using advanced 
	deep neural networks to analyze exospheric density measurements. The algorithm is adept at predicting relative
	elemental compositions and reconstructing surface maps, which is crucial for understanding how neutral atoms 
	pass from the planet’s regolith into its exosphere. Key characteristics of the method include the definition 
	and utilization of the simulated model's physical parameter space, the development of a sophisticated multilayer perceptron 
	DNN architecture, the application of Bayesian hyperparameter tuning for optimal configuration, and the integration of 
	domain-specific knowledge into the feature selection process.

	The principal accomplishment of this study is the creation of a DNN that accurately estimates surface-exosphere 
	interactions on Mercury. This algorithm effectively represents a specific region of the physical parameter 
	space, mapping complex relationships between exospheric and surface compositions. By yielding an 
	estimator for exosphere generation processes, the study bridges theoretical modeling with practical data-driven 
	predictions. Although the surface maps and exospheric model used in this analysis do not accurately depict 
	Mercury's real environment, this achievement has the potential to enhance our understanding of Mercury’s exosphere 
	and offers new avenues for research in planetary science and machine learning, suggesting potential for significant future discoveries.

	In fact, the primary aim of our paper is to validate the methodological framework, demonstrating that a 
	multilayer perceptron can infer an underlying compositional map from exospheric data under broad and 
	generalized conditions. To achieve this, we deliberately selected a set of simplified or not-yet-fully-realistic composition 
	models, ensuring that the machine learning pipeline remains adaptable and unbiased, rather than being constrained by a 
	predefined or overly specific representation of Mercury’s surface. Since this work is intended as a proof-of-concept, we did not set 
	out to reproduce the exact Mercury surface map. Instead, our simulated data help us assess whether the 
	network can handle large variability and still reliably derive compositional fractions. In future studies, we 
	plan to explore both expanding our process parameter space to determine how different regions are 
	represented by the algorithm and focusing on more realistic conditions to examine how the network predicts 
	physically faithful surface maps.

	We emphasize that this deliberately broad, less “true-to-reality” approach serves to validate our methodology 
	and does not preclude the adoption of more specific compositional models at a later stage. Should high-quality, 
	full-coverage exospheric measurements become available, the trained network can be re-applied or further 
	fine-tuned. Our current results thus constitute a foundational step, demonstrating the feasibility of 
	ML-based methods for this type of inverse problem.

\subsection{Performance and Results}

	A preliminary test phase demonstrated the model's superior predictive performance on Mercury's dayside 
	but highlighted less precise predictions for elements like sodium and potassium on the nightside. 
	These results indicate that some elemental species have a stronger link with the surface, retained even on
	the nightside (refractory elements), while others are more strongly connected to the surface on the dayside 
	(volatile elements) due to exospheric effects like radiation pressure, making nightside predictions
	less straightforward. Meanwhile, other species, such as oxygen and silicon, being widely present on the 
	surface, have less significance in the analysis. This led to a focused test campaign using only simulated daylight 
	observations from two consecutive Mercury years, leveraging the planet's spin-orbit resonance. This approach 
	significantly improved predictions for volatile species and enabled a comprehensive reconstruction of Mercury's 
	surface.

	The developed algorithms for surface composition reconstruction achieved high fidelity in their predictions, with 
	mean accuracy metrics reaching 89.70\% (ES4) and 83.41\% (R-squared) across 15 test sets at 200 km altitude. 
	Additionally, both mean absolute and relative residuals of elemental predictions showed a robust trend of decreasing 
	with altitude, reaching as low as 0.56\% and 14.70\% respectively. The effectiveness of these algorithms was 
	particularly notable in reconstructing elemental composition maps from low-altitude measurements (200km to ~700km),
	especially for sufficiently represented elements, such as aluminium, calcium, magnesium, and sodium. At the same time, 
	minor elements, such as water, exhibited jumps and anomalies in their predictions, attributed to their low overall 
	fraction in our model and their randomized placement on the surface (not constrained to the poles only). Meanwhile, 
	the MLP was able to capture the coupling of sulfur and zinc in its predictions, even if the input exospheric densities 
	are widely different for the two elements.

	We currently use ES4, R-squared, and absolute/relative residuals, finding these sufficient for a 
	multi-perspective evaluation in this proof-of-concept phase. However, we acknowledge that weighted metrics 
	or a deeper analysis of label distributions could reveal additional layers in how well the network prioritizes or predicts certain 
	elements under diverse data distributions. Additionally, our approach to generating large synthetic datasets and 
	training deep models can be computationally intensive, especially in its simulation and training phases (see Appendix C). Future 
	studies will aim to benchmark these resource demands, compare them with alternative frameworks, and 
	examine interpretability techniques (e.g., saliency maps or feature-attribution methods) to elucidate how the 
	DNN makes its predictions, and provide potential links between its internal structure and the physical parameter 
	space it represents.

	As a final remark, the present work reveals a promising new direction for investigating of surface-exosphere connections.
	Nevertheless, there are still many possible avenues for improvement before this tool can be considered fully mature. 
	The extensive training and testing campaigns conducted in this study highlight significant potential for refining the algorithms, while 
	the data generation models can be further improved with even more detailed and realistic representations of the physical processes. 
	In the following, we propose several future research paths to enhance our method, overcome its current limitations, and advance
	its applicability to real data.

\section{Future Work and Conclusions}

\subsection{Refining the Data Generation Model}

	While the MESSENGER space mission has significantly advanced our understanding of Mercury’s surface and exosphere, 
	our present study focuses on building the foundations of a method for training and studying data-driven representations
	of surface-exospheric interactions, rather than fine-tuning to specific datasets. 
	For this reason, we have maintained relatively broad parameter ranges to test performance under diverse assumptions. 
	In future research, we plan to refine these ranges and evaluate the model’s capability using MESSENGER-derived 
	or similar datasets—bridging the gap between method-development and real-data applications for upcoming 
	missions like BepiColombo.

	Future work includes expanding and refining the physical processes parameter space used in the data generation model. 
	This endeavor will align the algorithm more closely with the complex realities governing interactions 
	between a planet's surface and its exosphere. A key part of this exploration includes modeling previously 
	compressed dimensions within the parameter space, such as the complex chemistry within the MIV-produced cloud, 
	more complex impacting dust populations, more involved regolith effects on the release processes
	(e.g. diffusion), and potential interactions between the release processes, all aiming to more accurately reflect the intricate 
	dynamics of real-world processes. Likewise, parameters such as vibrational frequencies, binding energies, and sputtering yields 
	have been treated in simplified terms; subsequent analyses may integrate laboratory data for more accurate and variable
	process rates. Unraveling these dimensions, previously simplified in our model, is expected to provide 
	deeper insights into planetary science mechanisms and enhance the estimator’s ability to capture the full spectrum of 
	surface-exosphere interactions.

	Currently, our approach treats the surface composition as a closed set of elements, which can overlook the presence 
	of unknown or unmodeled species if they exist in significant quantities. In future work, we may explore adaptive output 
	layers or more flexible compositional assumptions. These would allow the model to remain robust even when confronted 
	with incomplete mineralogical information.
	Moreover, we aim to explore the performance of the DNNs on data distributions derived from more constrained 
	surface models that closely follow the observed mineralogy, elemental composition, and distribution of elements, 
	especially water, on the surface. In another avenue for future work we plan to investigate adaptive or finer-grained 
	tiling strategies to examine the capability of the algorithm to capture subtle spatial heterogeneities more effectively.

	In this study, we opted for a symmetric dipole representation, limiting dynamic coupling with the 
	interplanetary magnetic field. We recognize that Mercury’s offset dipole and variable magnetosphere 
	can substantially alter ion precipitation and, consequently, exospheric composition. Hence, introducing 
	a more realistic magnetic field configuration—including offset dipoles, time-dependent reconnection events, and interplanetary
	magnetic field effects— is a logical extension for future models.

\subsection{Exploring the Parameter Spaces}

	Another critical area of future research is testing DNNs trained with data from one region of the physical processes 
	parameter space against data distributions from different regions. This exploration is essential to assess the DNNs' 
	performance when applied to various models of exospheric production and their respective data distributions, 
	potentially constructing another layer for analyzing the surface-exosphere interactions with this innovative tool. 
	
	One potential application of this approach involves utilizing multiple DNNs trained on different regions of parameter 
	space. Input data could be passed through all of the pre-built DNNs, and their predictions juxtaposed in a subsequent 
	layer of the algorithm to estimate the input data's underlying generation mechanism. This would constrain the range 
	of physical parameters to those of the DNNs with the highest accuracy on the input data.

	Although we model the exosphere at a single point in time, ongoing processes—including solar-driven and 
	micrometeoroid-induced variations—may add some dynamism. Constructing a time-dependent parameter space, adapting 
	and training the DNN on temporal exospheric data would be an exciting extension, potentially 
	offering deeper insights into the relationship between transient events and Mercury’s surface composition.

	Another potentially groundbreaking utilization may explore the mapping between the physical processes space, 
	defined by analytical equations, and the purely data-driven DNN representation, built from the internal weight 
	matrices of the neural nets. This mapping could serve the dual purpose of understanding how a description of reality 
	constructed strictly from data relates to the description by physical equations, and exploring potential synergies between 
	the two in describing the real world.
	
	Additionally, expanding and elaborating on the hyperparameter space is identified as another area for development. 
	This will involve constructing a hyperparameter space that considers aspects such as network layer connectivity, 
	optimization of loss functions, and the functions used within the hidden and output units. Exploring alternative DNN 
	architectures also holds promise for enhancing the models' application, accuracy, and reliability. Further research 
	into feature engineering by applying more domain-specific knowledge to optimize input parameters can provide a 
	better representation of empirical data distributions. Observations of discrepancies between dayside and nightside 
	predictions may warrant an examination of split DNNs trained on data from only one side of the planet (illuminated 
	or shadowed). These developments aim to push the boundaries of what these algorithms can achieve in the analysis 
	of surface-exosphere interactions.

\subsection{Future Data and Real-World Validation}

	The contrast between using simulated data and incorporating real observational data into our algorithm development 
	merits further exploration. While simulations provide a controlled environment for testing various scenarios, they do 
	not capture the full complexity and unpredictability of actual exospheric data. This could lead to significant deviations 
	in the parameter space from those assumed in our simulations, as real processes and their interdependent variables 
	may change over time and are not fully represented in simulations.

	An important future step will be to test the network’s generalization on more comprehensive Mercury 
	datasets, including those from BepiColombo’s SERENA suite. This involves re-training, fine-tuning, or 
	combining multiple networks to accommodate diverse data distributions and, potentially, to identify which 
	physical processes dominate under various conditions. We believe this incremental path—from broad 
	synthetic coverage to increasingly realistic parameter spaces—will ensure that the method matures into a 
	fully applicable tool for exospheric analysis as robust observational data become available.

	Shifting our focus from simulated to real physical processes is a bold and potentially transformative step. Developing 
	an estimator capable of effectively processing and analyzing real-world data from Mercury's exosphere would 
	significantly advance our understanding of planetary surfaces and their interactions with their environments. This 
	progress would not only deepen our theoretical knowledge but also offer practical insights into the formation, dynamics, 
	and evolution of planetary exospheres.

	This study is performed in anticipation of the upcoming BepiColombo ESA/JAXA mission. The mission will deploy two 
	spacecraft—the Mercury Planetary Orbiter (MPO) and the Mercury Magnetospheric Orbiter (MMO)—equipped with a 
	suite of instruments aimed at understanding Mercury's surface, exosphere, and magnetosphere 
	\cite{benkhoff,milillo2,milillo3}. Particularly, our study targets future utilization of measurements from the SERENA 
	(Search for Exospheric Refilling and Emitted Natural Abundances) suite on the MPO, which includes a mass spectrometer – 
	STROFIO (STart from a ROtating Field mass spectrOmeter) –  for the characterization of the exosphere, an ion analyzer – 
	MIPA (Miniature Ion Precipitation Analyser) – and an ion spectrometer – PICAM (Planetary Ion CAMera) – for the 
	characterization of ion precipitation and the ionized environment, and an energetic neutral atom imager – ELENA 
	(Emitted Low Energy Neutral Atoms) – for mapping the ion precipitation regions via ENA \cite{orsini1,Orsini2021,milillo}. 
	Notable instruments aboard BepiColombo, which may provide images of the surface, include MIXS, MGNS, MERTIS, and 
	SIMBIO-SYS \cite{benkhoff}.

	The application of these methods to the observational data from BepiColombo's suite of instruments offers a promising 
	path to refine models of exosphere generation. By comparing predicted surface compositions with actual measurements, 
	we can more accurately constrain our models—thereby enhancing our understanding of planetary processes
	and, ultimately, informing broader theories of planetary exospheres.

\subsection{Conclusions and Outlook}

	The essence of our contribution lies in demonstrating a flexible ML approach for an inherently complex 
	inverse problem. Although we have not benchmarked it against classical methods or fully exploited the 
	breadth of Mercury’s known parameters, this initial framework provides a robust foundation for continued 
	development. As richer datasets become available and we expand our model’s physical fidelity, we anticipate 
	refining both the algorithm's architecture and its training protocols.

	Traditional analytical and Monte Carlo methods remain valuable, especially when the underlying physics is 
	well-characterized. However, several important Mercury processes are only partially understood or interlinked 
	in complex ways. In contrast, ML methods can accommodate large, high-dimensional datasets without relying on 
	strictly defined physical assumptions. By training on simulated data, we demonstrate that a neural network can 
	effectively invert exospheric observations, even if the planetary processes are intricate or not fully constrained. 
	As real data coverage expands, the ML approach can be re-applied or fine-tuned to reflect new insights on Mercury’s 
	exosphere.

	Additionally, investigating the data-driven representations themselves—particularly the algorithm’s ability to build an 
	estimator of broad regions in parameter space—could be of significant interest. Finally, a thorough comparison of results 
	obtained by traditional models and those generated by data-driven ML methods (once trained on real data) is crucial 
	for examining the release processes in depth.

	In conclusion, this research establishes a solid foundation for advancing our understanding of planetary 
	surface-exosphere interactions, particularly around Mercury. By utilizing virtual exospheric measurements as inputs to 
	deep neural networks, we have taken a significant step forward, enhancing the capabilities of estimators and broadening 
	our understanding of planetary science. The application of this method to the anticipated data collected by the 
	BepiColombo mission will represent a notable advance in space exploration. With sophisticated AI algorithms, 
	BepiColombo's potential to uncover insights into Mercury's exosphere dynamics will be greatly enhanced. Moreover, 
	the ongoing development and refinement of deep neural networks in this study promise to revolutionize our approach 
	to studying planetary bodies within our Solar System, providing new tools for understanding the complex processes 
	that govern the environments of celestial objects.

\appendix
\section{Detailed Description of the MLP DNN Architecture}

\subsection{Deep Neural Network Architecture}

	The MLP, a class of feedforward neural network, excels in multivariate regression by modeling complex 
	nonlinear functions with its multi-layered, fully connected structure and nonlinear activation functions 
	\cite{minsky, rumelhart1, kingma, lecun}. This architecture, combined with optimization techniques like 
	backpropagation, allows MLPs to identify intricate patterns in high-dimensional data, making them ideal for 
	robust predictive modeling.

	Training involves preprocessing data for network suitability, building the model using the Keras framework 
	with TensorFlow \cite{tensorflow2015, chollet2015}, and iteratively tuning the network through backpropagation 
	to minimize error \cite{rumelhart2}. This process ensures effective and reproducible model performance.

	The architecture of a multilayer perceptron enables complex data processing through a structured 
	network of layers: an input layer, multiple hidden layers for nonlinear transformations, and an output layer 
	for predictions. The network's effectiveness hinges on key components like the loss function, which guides 
	accuracy improvements, and the regularizer, which ensures generalizability. Efficiently chosen optimization 
	algorithms and precise hyperparameter tuning further enhance the network's performance. Figure 
	\ref{fig:MLP-arch} illustrates this interplay, crucial for tasks like analyzing Mercury's exosphere, with 
	subsequent sections detailing each component's role in predictive capabilities. The inner connectiveness 
	of the MLP DNN neural units is shown on Figure \ref{fig:MLP-est-basic}.

	\subparagraph{Input Layer}\hfill

	The input layer of the MLP introduces data, in our case Mercury’s exospheric density measurements, 
	into the network, with each neuron representing a distinct data feature. For example, distinct elemental 
	density measurements are represented by separate neurons. Before entry, data undergo normalization to 
	ensure uniform influence on the learning process, thereby preventing bias \cite{goodfellow1}. This involves 
	standardizing each feature to zero mean and unit variance as per the equation:

	\begin{equation}
		\boldsymbol{x} = \frac{\boldsymbol{x_{\text{orig}}} - \boldsymbol{\mu}}{\boldsymbol{\sigma}}
	\end{equation}

	where $\boldsymbol{x}$ is the standardized vector of input features, $\boldsymbol{x_{\text{orig}}}$ is the 
	original vector of input features, $\boldsymbol{\mu}$ is the vector of means of the feature values, and 
	$\boldsymbol{\sigma}$ is the standard deviations vector. Such standardization enhances the efficiency and 
	stability of the network's learning process.

	\subparagraph{Hidden Layers}\hfill

	The hidden layers form the core of the MLP architecture, where the actual processing and learning 
	occur \cite{minsky, hinton1}. Positioned between the input and output layers, they transform 
	input data into a form usable for predictions. Each hidden layer is composed of a set of neural units - 
	neurons - and each neuron in these layers is fully connected to all neurons in the preceding and 
	succeeding layers, creating a dense network of synaptic connections. The structure of these layers is represented 
	mathematically by combination matrices or weight matrices, which, along with the activation function 
	applied at each neuron, helps form an estimation of the relationships among the processes acting between 
	the input layer and the output layer.

	In our study on Mercury's exosphere, multiple hidden layers with a substantial number of neurons allow 
	the MLP to capture the nuances of Mercury's exospheric composition and the underlying processes that govern it. 
	A key component of these hidden layers is the activation function, in this case the Rectified Linear Unit (ReLU), 
	essential for introducing nonlinearity and aiding in effective gradient propagation to avoid vanishing gradients 
	\cite{glorot}. The ReLU function is defined as $a(\boldsymbol{z}) = \textrm{max}(0, \boldsymbol{z})$, 
	where $\boldsymbol{z}$ is the input to the activation function.

	The transformation within each hidden layer then follows the equation:

	\begin{equation}
		\boldsymbol{h} = a(\boldsymbol{z}) = a(\boldsymbol{W}^T\boldsymbol{x} + \boldsymbol{b}) = max(0, \boldsymbol{W}^T\boldsymbol{x} + \boldsymbol{b}),
	\end{equation}

	where $\boldsymbol{W}^T$ represents the weight matrix, $\boldsymbol{x}$ is the input vector to the
	hidden layer (input features or activations from a previous hidden layer), and $\boldsymbol{b}$ is the
	bias vector of the affine transformation. This equation encapsulates the affine transformation followed 
	by the application of the ReLU activation function, enabling the network to learn and represent complex 
	nonlinear relationships.

	Finally, the output from the hidden layers is passed on to the output layer, where the final prediction is made. 
	The architecture and depth of the hidden layers are critical and typically determined through empirical methods 
	and hyperparameter tuning. This ensures the network has the requisite complexity for effective learning while 
	avoiding overfitting to the empirical distribution present in the training data.

	\subparagraph{Output Layer}\hfill
	
	The output layer is the final layer in an MLP, playing the role of determining the format and nature of its 
	predictions. In the context of our study, this layer is tailored to predict the elemental composition of Mercury's 
	surface, with each neuron corresponding to one of the elements being analyzed. For example, if predicting 
	the fractions of 11 different elements, the output layer would consist of 11 neurons.

	The activation function used in the output layer is crucial and depends on the nature of the prediction task. 
	In our case, where the output is a set of continuous values that sum to 1 (representing 
	fractions), the softmax function is used \cite{joachims}. The softmax function converts the 
	raw output of the network into a probability distribution, ensuring that the predicted fractions are non-negative 
	and sum up to one, aligning perfectly with the physical reality of our task.
	
	The formula for the softmax function is as follows:
	
	\begin{equation}
		\hat{y_i} = \textrm{softmax}(\boldsymbol{z})i = \frac{\textrm{exp}(z_i)}{\sum_{j}{\textrm{exp}(z_j)}},
	\end{equation}
	
	where $\hat{y_i}$ is the predicted fraction for the $i$-th element, and $z$ represents the raw output values from 
	the final hidden layer. This configuration allows the network to deliver accurate, meaningful predictions of 
	Mercury’s surface composition, synthesizing the representation insights gained from all previous layers.

\subsection{Loss Function, Regularization, and Optimization}

	\subparagraph{Loss Function}\hfill

	The loss function plays a pivotal role in guiding the optimization process, quantifying the discrepancy 
	between the network's predictions and the actual target values to gauge model accuracy. In  our most 
	successful MLP tests on predicting Mercury's surface elemental composition, the Kullback-Leibler (KL) 
	divergence \cite{cover} has proven particularly effective. It measures how one probability distribution, 
	representing the predicted elemental composition (the output from the MLP), diverges from the actual 
	distribution (the true elemental composition). The formula for KL divergence is:

	\begin{equation}
		KL(P || Q) = J(\boldsymbol{W, b, x, y}) = \sum_{i} P(i) \log\frac{P(i)}{Q(i)},
	\end{equation}
	
	where $P$ represents the true distribution of the fraction of element $i$ in the data, and $Q$ is the predicted 
	distribution from the MLP. Other loss functions like Categorical Cross Entropy, Mean Absolute Error (MAE), and 
	Mean Squared Error (MSE) were also considered. However, the KL divergence was preferred for our regression 
	task because it aligns better with the probabilistic requirements, focusing on relative proportions rather than 
	absolute quantities of elements. At the same time, it showed marginally better performance in our initial tests
	than similarly probabilistic in nature loss functions, such as Categorical Cross Entropy.

	We should note here that in our Keras-based implementation we rely on the library’s default numerical stabilization 
	for KL divergence, which handles extremely small or zero values to avoid undefined operations. In future extensions, 
	we may explore a custom offset or other approaches if more granular control over zero probabilities becomes necessary,
	especially when we consider potential expansions of the available pool of surface elements.

	\subparagraph{Regularization}\hfill

	Regularization is an essential technique in neural network training, designed to enhance model generalization 
	by adding constraints or penalties to the loss function. In our study, we use L2 regularization (weight decay) 
	on the weights of each hidden layer\cite{bishop2}. This technique constrains the magnitude of the weights, 
	preventing them from becoming excessively large and helping to avoid overfitting the model to the specific 
	dataset used for training.

	The L2 regularization is mathematically represented as:

	\begin{equation}
		\hat{J}(\boldsymbol{W, b, x, y}) = J(\boldsymbol{W, b, x, y}) + \lambda\sum_{i=1}^{m}|\boldsymbol{\theta}_i|^2,
	\end{equation}

	where $J(\boldsymbol{W, b, x, y})$ is the original loss function, $\lambda$ is the regularization coefficient, 
	and $\theta$ denotes the vector of all weight parameters (unfolded from the $\boldsymbol{W}$ matrices). 
	The right choice of $\lambda$ is critical. If $\lambda$ is too large, it can lead to underfitting, where the model 
	is overly simplified and fails to capture the underlying trends in the data. Conversely, a very small $\lambda$ 
	might not effectively prevent overfitting. 

	In a multivariate regression task such as ours, where the model needs to understand complex relationships 
	between various features in the surface-exosphere interaction at Mercury, L2 regularization helps in 
	maintaining a balance between MLP model complexity and its ability to generalize. The addition of this 
	regularization term (penalty) to the loss function thus ensures that the model not only fits the training data 
	well but also maintains the flexibility to perform accurately on new, unseen data.

	\subparagraph{Optimization}\hfill

	The training of our multilayer perceptron for predicting Mercury's surface composition employs the Adam 
	optimization algorithm, a refinement of stochastic gradient descent known for its effectiveness with 
	large-scale data and complex models \cite{kingma}. The fundamental mechanism of Adam involves 
	dynamically and adaptively updating the weights of the combination matrices for each hidden layer 
	to minimize the total error as indicated by the loss function. This is achieved through backpropagation 
	optimization \cite{rumelhart1, rumelhart2}, where the weights are adjusted following their gradients 
	with respect to the loss function:

	\begin{equation}
		\boldsymbol{\theta} := \boldsymbol{\theta} - \alpha \frac{1}{m} \bigtriangledown_{\boldsymbol{\theta}} \text{KL}(P|Q) = \boldsymbol{\theta} - \alpha \frac{1}{m} \bigtriangledown_{\boldsymbol{\theta}} \sum_{j=1}^{m} \sum_{i} P_j(i) \log \frac{P_j(i)}{Q_j(i)},
	\end{equation}

	In this equation, $\alpha$ represents the learning rate and $\bigtriangledown_{\boldsymbol{\theta}} \text{KL}(P|Q)$ 
	is the gradient of the KL divergence with respect to the model parameters $\theta$. The stochastic nature of 
	the gradient descent implies that learning iterations are not performed on the entire dataset but rather on a 
	random subset known as a mini-batch. Here, $m$ denotes the number of examples in the mini-batch.

\subsection{Hyperparameter Tuning}\hfill

	Key hyperparameters of our MLP DNN are:

	\begin{itemize}
		\item \textbf{Learning Rate}: This parameter governs the size of the steps taken during the backpropagation optimization 
		algorithm along the weight gradients of the loss function. A well-balanced learning rate is critical—it must be large 
		enough to navigate plateaus in the loss function's parameter space, yet sufficiently small to converge to (or remain near) 
		the minimum of the error.
	
		\item \textbf{Mini-Batch Size}: This refers to the size of the random subset of examples used in each training iteration, 
		impacting both the speed and stability of the learning process.
	
		\item \textbf{Number of Hidden Layers and Neurons}: These parameters determine the depth and width of the neural 
		network, influencing its ability to model complex relationships in the data.

		\item \textbf{L2 Regularization Coefficient}: This defines the degree of penalty imposed on large weight values, helping 
		to prevent overfitting by controlling model complexity.
	\end{itemize}

	To fine-tune these hyperparameters, we employed a Bayesian optimization strategy using the Gaussian Process (GP) 
	approach, as outlined in \citeA{bergstra}. The tuning process was facilitated by the scikit-optimize library 
	\cite{scikit-opt}, which utilizes a prior probability distribution function to identify the hyperparameter configuration 
	that minimizes the total loss on a hold-out validation dataset. This systematic adjustment of hyperparameters not 
	only enhances learning capabilities and overall performance, but also optimizes the balance between model 
	complexity and efficiency.

\subsection{Performance Metrics}\hfill

	To evaluate the performance of our machine learning model, we utilize both customized and standard metrics 
	to ensure precise and insightful quantitative assessments. Our primary metric, the Euclidean 
	similarity 4 (ES4), integrates elements of Euclidean distance and cosine similarity, providing a nuanced measure 
	of prediction accuracy by considering both magnitude and directionality in multidimensional space:

	\begin{equation}
		\textrm{ES4} = \left(1 - \frac{\sqrt{\sum_{i}(\boldsymbol{\hat{y}_i} - \boldsymbol{y_i})^2}}{\sqrt{\sum_{i}\boldsymbol{y_i}^2}}\right) \times \left(\frac{\boldsymbol{\hat{y}_i} \cdot \boldsymbol{y_i}}{\|\boldsymbol{\hat{y}_i}\| \|\boldsymbol{y_i}\|}\right),
	\end{equation}

	where $\boldsymbol{\hat{y_i}}$ and $\boldsymbol{y_i}$ represent the predicted and actual surface compositions, 
	respectively.

	Moreover, we incorporate the R-squared ($R^2$) metric into our evaluation framework. The $R^2$ metric, 
	commonly used in regression analysis, quantifies the proportion of the variance in the dependent variable that is 
	predictable from the independent variable(s). In the context of multivariate regression, $R^2$ is defined as:

	\begin{equation}
		R^2 = 1 - \frac{\sum_{i}(\boldsymbol{y_i} - \boldsymbol{\hat{y_i}})^2}{\sum_{i}(\boldsymbol{y_i} - \bar{\boldsymbol{y}})^2},
	\end{equation}

	where $\bar{\boldsymbol{y}}$ is the mean of the actual values. This metric is particularly useful for assessing 
	the model's ability to capture the variance in the data, offering insights into how well the model's predictions 
	approximate the actual data distribution compared to a naive model that only predicts the mean.

	Additionally, we evaluate the model using absolute and relative residuals, which provide further granularity in 
	understanding the model's performance. These residuals help identify the absolute and relative differences between 
	predicted and actual values, offering a direct measure of prediction error.

	By combining these metrics, we achieve a multidimensional evaluation of our DNN's performance, encompassing 
	both the accuracy of individual predictions and the model's overall ability to capture the complexity of the data. 
	This comprehensive assessment not only ensures validation of the model's outputs but also sheds light on areas 
	for potential improvement, thereby contributing to the refinement of the model's predictive capabilities.

\clearpage
\section{Detailed Datasets}

\begin{table}[h]
    \centering
    \caption{Mean surface mineral fractions in the datasets used for training, validation, and testing.}
  	\label{tab:appa_mins}
    \resizebox{\textwidth}{!}{%
    \begin{tabular}{l c c c c c c c c c}
        \toprule
        \textbf{Surfaces} & \textbf{Albite} & \textbf{Anorthite} & \textbf{Diopside} & \textbf{Enstatite} & \textbf{Ferrosilite} & \textbf{Hedenbergite} & \textbf{Orthoclase} & \textbf{Sphalerite} & \textbf{Water Ice} \\
        \midrule
        Training x10  & 0.102 & 0.170 & 0.146 & 0.156 & 0.116 & 0.079 & 0.110 & 0.089 & 0.032 \\
        Training x20  & 0.109 & 0.136 & 0.162 & 0.149 & 0.126 & 0.072 & 0.131 & 0.077 & 0.038 \\
        Training x40  & 0.136 & 0.124 & 0.148 & 0.142 & 0.133 & 0.065 & 0.139 & 0.065 & 0.046 \\
        Training x60  & 0.133 & 0.124 & 0.153 & 0.137 & 0.152 & 0.062 & 0.137 & 0.056 & 0.046 \\
        Training x80  & 0.134 & 0.121 & 0.147 & 0.134 & 0.154 & 0.066 & 0.138 & 0.060 & 0.046 \\
        Training x100 & 0.131 & 0.127 & 0.144 & 0.134 & 0.151 & 0.070 & 0.135 & 0.058 & 0.048 \\
        Training x150 & 0.133 & 0.133 & 0.143 & 0.139 & 0.137 & 0.066 & 0.137 & 0.067 & 0.047 \\
        Training x200 & 0.139 & 0.134 & 0.143 & 0.136 & 0.133 & 0.065 & 0.133 & 0.072 & 0.045 \\
        Training x300 & 0.140 & 0.134 & 0.141 & 0.137 & 0.137 & 0.065 & 0.134 & 0.069 & 0.044 \\
        \midrule
        v01  & 0.042 & 0.310 & 0.088 & 0.119 & 0.103 & 0.061 & 0.101 & 0.062 & 0.114 \\
        t01  & 0.095 & 0.116 & 0.090 & 0.120 & 0.397 & 0.017 & 0.107 & 0.020 & 0.038 \\
        t02  & 0.384 & 0.124 & 0.131 & 0.098 & 0.053 & 0.020 & 0.138 & 0.032 & 0.021 \\
        t03  & 0.099 & 0.150 & 0.202 & 0.124 & 0.085 & 0.114 & 0.124 & 0.053 & 0.050 \\
        t04  & 0.148 & 0.182 & 0.173 & 0.132 & 0.088 & 0.069 & 0.139 & 0.034 & 0.034 \\
        t05  & 0.125 & 0.054 & 0.063 & 0.027 & 0.325 & 0.020 & 0.087 & 0.241 & 0.058 \\
        t06  & 0.097 & 0.113 & 0.095 & 0.153 & 0.095 & 0.140 & 0.116 & 0.149 & 0.043 \\
        t07  & 0.241 & 0.075 & 0.191 & 0.144 & 0.200 & 0.011 & 0.089 & 0.033 & 0.017 \\
        t08  & 0.074 & 0.146 & 0.064 & 0.073 & 0.219 & 0.153 & 0.178 & 0.051 & 0.043 \\
        t09  & 0.151 & 0.146 & 0.127 & 0.063 & 0.086 & 0.019 & 0.338 & 0.025 & 0.045 \\
        t10  & 0.285 & 0.176 & 0.196 & 0.182 & 0.056 & 0.022 & 0.047 & 0.020 & 0.017 \\
        t11  & 0.261 & 0.222 & 0.104 & 0.242 & 0.047 & 0.012 & 0.085 & 0.007 & 0.021 \\
        t12  & 0.172 & 0.237 & 0.187 & 0.208 & 0.086 & 0.010 & 0.074 & 0.008 & 0.018 \\
        t13  & 0.295 & 0.259 & 0.140 & 0.125 & 0.076 & 0.018 & 0.065 & 0.009 & 0.013 \\
        t14  & 0.178 & 0.240 & 0.131 & 0.157 & 0.040 & 0.008 & 0.166 & 0.069 & 0.011 \\
        t15  & 0.186 & 0.265 & 0.145 & 0.109 & 0.105 & 0.010 & 0.102 & 0.068 & 0.010 \\
        \bottomrule
    \end{tabular}
    }
\end{table}

We note that we use stoichiometric (atomic) fractions rather than mass or volume fractions 
when converting mineral abundances to elemental abundances. Specifically, each mineral’s atomic ratio is 
multiplied by the percentage of that mineral in the mixture, and the results are summed across all minerals 
to obtain the total atomic fraction for each element. This may lead to slight discrepancies when comparing 
the fraction of a single mineral to its final elemental contribution, due to additional elements contributed by 
other minerals.

\begin{table}[h]
    \centering
    \caption{Mean surface elemental fractions in the datasets used for training, validation, and testing.}
  	\label{tab:appa_elems}
    \resizebox{\textwidth}{!}{%
    \begin{tabular}{l c c c c c c c c c c c c}
        \toprule
        \textbf{Surfaces} & \textbf{Na} & \textbf{Al} & \textbf{Si} & \textbf{O$_2$} & \textbf{Ca} & \textbf{Mg} & \textbf{Fe} & \textbf{K} & \textbf{Zn} & \textbf{S} & \textbf{H$_2$O} \\
        \midrule
        Training x10  & 0.014 & 0.074 & 0.275 & 0.421 & 0.055 & 0.066 & 0.045 & 0.015 & 0.015 & 0.015 & 0.005 \\
        Training x20  & 0.015 & 0.069 & 0.279 & 0.421 & 0.052 & 0.066 & 0.047 & 0.018 & 0.014 & 0.014 & 0.007 \\
        Training x40  & 0.018 & 0.069 & 0.282 & 0.422 & 0.047 & 0.062 & 0.048 & 0.019 & 0.011 & 0.011 & 0.010 \\
        Training x60  & 0.017 & 0.068 & 0.284 & 0.424 & 0.047 & 0.061 & 0.052 & 0.018 & 0.009 & 0.009 & 0.009 \\
        Training x80  & 0.018 & 0.068 & 0.284 & 0.424 & 0.046 & 0.059 & 0.053 & 0.018 & 0.010 & 0.010 & 0.009 \\
        Training x100 & 0.017 & 0.069 & 0.283 & 0.424 & 0.047 & 0.059 & 0.053 & 0.019 & 0.010 & 0.010 & 0.009 \\
        Training x150 & 0.018 & 0.071 & 0.282 & 0.423 & 0.048 & 0.060 & 0.049 & 0.018 & 0.012 & 0.012 & 0.008 \\
        Training x200 & 0.018 & 0.072 & 0.282 & 0.422 & 0.048 & 0.060 & 0.047 & 0.018 & 0.012 & 0.012 & 0.008 \\
        Training x300 & 0.019 & 0.072 & 0.282 & 0.423 & 0.047 & 0.059 & 0.048 & 0.018 & 0.012 & 0.012 & 0.008 \\
        \midrule
        v01  & 0.006 & 0.109 & 0.257 & 0.421 & 0.067 & 0.047 & 0.039 & 0.014 & 0.010 & 0.010 & 0.021 \\
        t01  & 0.012 & 0.058 & 0.285 & 0.430 & 0.031 & 0.046 & 0.113 & 0.014 & 0.003 & 0.003 & 0.006 \\
        t02  & 0.048 & 0.095 & 0.302 & 0.435 & 0.034 & 0.042 & 0.016 & 0.017 & 0.004 & 0.004 & 0.003 \\
        t03  & 0.013 & 0.071 & 0.280 & 0.426 & 0.066 & 0.064 & 0.041 & 0.016 & 0.008 & 0.008 & 0.007 \\
        t04  & 0.019 & 0.085 & 0.284 & 0.431 & 0.056 & 0.059 & 0.034 & 0.018 & 0.005 & 0.005 & 0.005 \\
        t05  & 0.020 & 0.053 & 0.263 & 0.386 & 0.023 & 0.019 & 0.106 & 0.016 & 0.052 & 0.052 & 0.012 \\
        t06  & 0.014 & 0.062 & 0.269 & 0.405 & 0.054 & 0.058 & 0.051 & 0.016 & 0.032 & 0.032 & 0.008 \\
        t07  & 0.031 & 0.062 & 0.295 & 0.431 & 0.038 & 0.064 & 0.056 & 0.011 & 0.005 & 0.005 & 0.003 \\
        t08  & 0.010 & 0.069 & 0.283 & 0.426 & 0.050 & 0.030 & 0.087 & 0.023 & 0.008 & 0.008 & 0.007 \\
        t09  & 0.019 & 0.098 & 0.298 & 0.435 & 0.037 & 0.033 & 0.025 & 0.043 & 0.003 & 0.003 & 0.006 \\
        t10  & 0.035 & 0.084 & 0.289 & 0.434 & 0.051 & 0.051 & 0.017 & 0.016 & 0.003 & 0.003 & 0.002 \\
        t11  & 0.030 & 0.097 & 0.286 & 0.436 & 0.043 & 0.078 & 0.014 & 0.011 & 0.001 & 0.001 & 0.003 \\
        t12  & 0.020 & 0.090 & 0.280 & 0.435 & 0.056 & 0.081 & 0.024 & 0.009 & 0.001 & 0.001 & 0.002 \\
        t13  & 0.034 & 0.106 & 0.284 & 0.438 & 0.052 & 0.051 & 0.022 & 0.008 & 0.001 & 0.001 & 0.002 \\
        t14  & 0.023 & 0.105 & 0.281 & 0.431 & 0.049 & 0.059 & 0.012 & 0.021 & 0.009 & 0.009 & 0.002 \\
        t15  & 0.024 & 0.105 & 0.277 & 0.431 & 0.055 & 0.048 & 0.029 & 0.013 & 0.009 & 0.009 & 0.001 \\
        \bottomrule
    \end{tabular}
    }
\end{table}

\begin{sidewaystable}
    \centering
    \caption{Test and validation datasets prepared for the generalization evaluation of the MPL DNN algorithm. 
	Each test surface-exosphere pair gives rise to one dataset per altitude level and per Mercury TAA. The
	"0 and 360" TAA signifies that two simulations are performed on this surface coming from two consequtive
	perihelia.}
  	\label{tab:test_sets}
    \resizebox{\textwidth}{!}{%
    \begin{tabular}{l l c c c c c}
        \toprule
        \textbf{Surf-Exo Pair Name} & \textbf{Resulting Dataset Type} & \textbf{\# of Examples Per Dataset} & \textbf{Mercury TAA} & \multicolumn{2}{c}{\textbf{Altitude Range [km]}} & \textbf{Total \# of Datasets from Surf-Exo pair} \\
        \midrule
        v01  & Validation & 648 & 0 & 500 & - & 1 \\
        t01  & Test & 648 & 0 and 360 & 200, 300, 400, 500, 600, 700, 800, 1000, 1500 & - & 18 \\
        t02  & Test & 648 & 0 and 360 & 200, 300, 400, 500, 600, 700, 800, 1000, 1500 & - & 18 \\
        t03  & Test & 648 & 0 and 360 & 200, 300, 400, 500, 600, 700, 800, 1000, 1500 & - & 18 \\
        t04  & Test & 648 & 0 and 360 & 200, 300, 400, 500, 600, 700, 800, 1000, 1500 & - & 18 \\
        t05  & Test & 648 & 0 and 360 & 200, 300, 400, 500, 600, 700, 800, 1000, 1500 & - & 18 \\
        t06  & Test & 648 & 0 and 360 & 200, 300, 400, 500, 600, 700, 800, 1000, 1500 & - & 18 \\
        t07  & Test & 648 & 0 and 360 & 200, 300, 400, 500, 600, 700, 800, 1000, 1500 & - & 18 \\
        t08  & Test & 648 & 0 and 360 & 200, 300, 400, 500, 600, 700, 800, 1000, 1500 & - & 18 \\
        t09  & Test & 648 & 0 and 360 & 200, 300, 400, 500, 600, 700, 800, 1000, 1500 & - & 18 \\
        t10  & Test & 648 & 0 and 360 & 200, 300, 400, 500, 600, 700, 800, 1000, 1500 & - & 18 \\
        t11  & Test & 648 & 0 and 360 & 200, 300, 400, 500, 600, 700, 800, 1000, 1500 & - & 18 \\
        t12  & Test & 648 & 0 and 360 & 200, 300, 400, 500, 600, 700, 800, 1000, 1500 & - & 18 \\
        t13  & Test & 648 & 0 and 360 & 200, 300, 400, 500, 600, 700, 800, 1000, 1500 & - & 18 \\
        t14  & Test & 648 & 0 and 360 & 200, 300, 400, 500, 600, 700, 800, 1000, 1500 & - & 18 \\
        t15  & Test & 648 & 0 and 360 & 200, 300, 400, 500, 600, 700, 800, 1000, 1500 & - & 18 \\
        \bottomrule
    \end{tabular}
    }
\end{sidewaystable}

\begin{table}[h]
    \centering
    \caption{Main training datasets used to train the MLP DNN algorithm. Each training dataset defines a
	different empirical data distribution which its respective MLP DNN is trained to approximate.}
  	\label{tab:train_sets}
	\resizebox{\textwidth}{!}{%
    \begin{tabular}{c l c c c}
        \toprule
        \textbf{\# Subsets} & \textbf{Data Augmentation} & \textbf{Examples per Subset} & \textbf{Total Examples} & \textbf{Altitude Range [km]} \\
        \midrule
        10  & No  & 648  & 6,480    & 500 \\
        20  & No  & 648  & 12,960   & 500 \\
        40  & No  & 648  & 25,920   & 500 \\
        60  & No  & 648  & 38,880   & 500 \\
        80  & No  & 648  & 51,840   & 500 \\
        100 & No  & 648  & 64,800   & 500 \\
        150 & No  & 648  & 97,200   & 500 \\
        200 & No  & 648  & 129,600  & 500 \\
        300 & No  & 648  & 194,400  & 500 \\
        200 & Yes & 4,536 & 907,200  & 200-2000 \\
        300 & Yes & 4,536 & 1,360,800 & 200-2000 \\
        \bottomrule
    \end{tabular}
	}
\end{table}

\clearpage
\section{Detailed Training Campaign}

\subparagraph{Starting Point}\hfill
	
	Our journey in the big training campaign endevoir commenced with a baseline training dataset, 
	which, being the simplest, was least representative of the true data generation distribution. 
	This dataset comprised 10 subsets with a total of 6,480 data points from the F00 feature set 
	(only exospheric density measurements). The initial architecture of the MLP DNN was 
	modeled closely after the structure proposed in the preliminary study of \citeA{kazakov}, featuring a four-layered 
	network with 400, 200, 200, and 100 neurons, respectively. Notably, our study expanded the input 
	layer to accommodate a greater number of elements - 11 total input elements.

	The output layer of the network employs softmax units, designed to predict the surface composition 
	of the same 11 elements provided as inputs. Initially, the minibatch size was set to 1,024 examples. 
	The regularization L-2 coefficient and the learning rate were chosen as $1.0\times10^{-6}$ 
	and $0.5\times10^{-4}$, respectively, to balance the trade-off between learning efficiency and the risk of overfitting.

\subparagraph{Eliminating Skewed Predictions}\hfill

	The initial analysis of predicting fractionated surface elemental composition revealed that the accuracy
	metrics were significantly skewed by the prevalence of certain abundant elements, notably oxygen (O$_2$)
	and silicon (Si), which are omnipresent in most of the minerals in our model. This skewness, stemming 
	from the algorithm's propensity to more accurately predict these
	two elements, was addressed by excluding them from the prediction vector and adjusting it to ensure a
	normalized sum of 1. Consequently, the refined model focuses 
	on predicting the normalized proportions of the remaining nine elements, with a subsequent 
	denormalization process applied for the map reconstruction purposes. This strategic exclusion of the 
	most abundant elements led to a marked enhancement of approximately 4\% in the predictive R$^2$ 
	accuracy for the other nine elements. It is important to note, however, that the input vector maintained 
	its original composition of 11 elements.

	This decision to modify the output layer by removing two elements was driven by a clear rationale: 
	the omnipresent elements, though significant, held less interest for the objectives of our study compared 
	to the other elements. This approach underscores our commitment to optimizing the model's performance 
	where it matters most, despite recognizing that alternative configurations of the output layer might exist.

\subparagraph{Training Set Size and Learning Curve Examination}\hfill

	Exploring the behavior of the initial MLP DNN involved examinining its performance in relation 
	to the expansion of the training dataset size and the extension of training duration. The aim 
	was to demonstrate the algorithm's nominal operation during both training and inference phases 
	by analyzing its learning curves. This included assessing training and generalization accuracies 
	across a training dataset and the hold-out validation dataset, respectively. Additionally, identifying 
	the optimal training duration for inference was crucial to mitigate the risk of overfitting, in line 
	with the guidance provided by \citeA{bengio}.

	Our investigation spanned training sets ranging from 10 to 200 unaugmented data subsets. We 
	observed a clear positive relationship between increasing the dataset size and the enhancement 
	of generalization accuracy. 

	In parallel, the algorithm's behavior was monitored in terms of its optimization process over 
	multiple iterations (epochs) across the entire training dataset, employing stochastic gradient 
	descent to converge to the minimum of the loss function. Analysis of the learning curves revealed 
	a maximum in predictive performance on the validation dataset after 40 epochs. This was in  
	contrast to the outcomes observed at 200 epochs of SGD, despite the training dataset's distribution 
	increasingly aligning with each additional training iteration.

\begin{figure}[h!]
		\centering
		\includegraphics[scale=0.7]{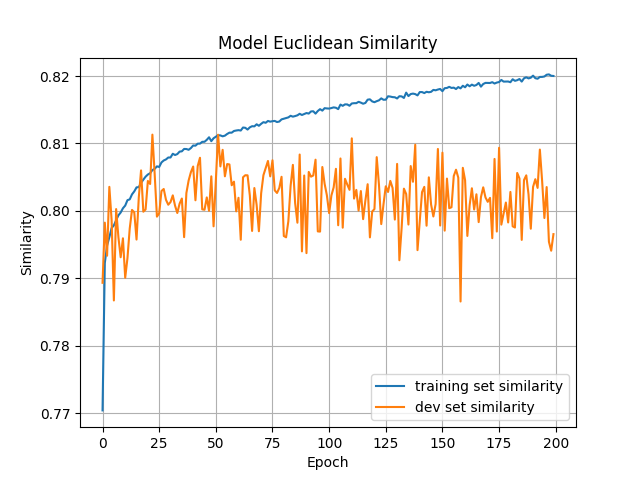}
		\caption{Learning curve for the MLP DNN training. The blue and orange curves show the evolution 
		of the average prediction similarity of the full training dataset 
		(200 subsets, 129,600 data points) and the development hold-out validation dataset (1 subset, 
		648 data points) respectively.
		}
  			\label{fig:learn_curve_200}
	\end{figure}

	The learning curve depicted in Figure \ref{fig:learn_curve_200} also hints at the potential for further 
	enhancements in training predictions, given the rapid ascent observed towards the training's 
	culmination. However, to ensure robust inference capabilities, it's imperative to diminish the 
	variance. This could potentially be achieved by incorporating a greater number of training examples 
	and/or intensifying the regularization measures.

\subparagraph{Feature Sets Examination}\hfill

	The evaluation of feature sets played a pivotal role in optimizing the performance of the MLP 
	DNN, particularly through the training of the network with various input features 
	across the unaugmented 200-subset training dataset. The assortment of the examined feature sets 
	is detailed in Table \ref{tab:feat_sets}.
	The iterative process of enhancing input features resulted in substantial improvements in 
	prediction accuracy for several modified feature sets with ES4 going up by about 2\%, while
	R$^2$ increasing with as much as 5\% from F00 to F11. This underscores the 
	critical importance of a well-curated and comprehensive feature set in the development of neural 
	networks capable of tackling intricate tasks, such as predicting the surface elemental compositions.

	\begin{table}[h]
	    \centering
	    \caption{Feature sets used in training and testing of the MLP DNN algorithm.}
  		\label{tab:feat_sets}
		\resizebox{\textwidth}{!}{%
	    \begin{tabular}{l l l l l l l}
	        \toprule
	        \textbf{Feature Set Name} & \textbf{Exospheric Densities} & \textbf{Altitude} & \textbf{Longitude} & \textbf{Latitude} & \textbf{Local Solar Time} & \textbf{Ion Precipitation} \\
	        \midrule
	        F00 & linear      & -           & -    & -    & -      & -      \\
	        F01 & linear      & linear      & -    & -    & -      & -      \\
	        F02 & logarithmic & -           & -    & -    & -      & -      \\
	        F03 & logarithmic & -           & -    & -    & -      & -      \\
	        F04 & logarithmic & -           & cos  & sin  & -      & -      \\
	        F05 & logarithmic & linear      & cos  & sin  & -      & -      \\
	        F06 & logarithmic & logarithmic & cos  & sin  & -      & -      \\
	        F08 & logarithmic & linear      & -    & -    & linear & linear \\
	        F09 & logarithmic & -           & cos  & sin  & linear & linear \\
	        F10 & logarithmic & linear      & cos  & sin  & linear & linear \\
	        F11 & logarithmic & logarithmic & cos  & sin  & linear & linear \\
	        \bottomrule
	    \end{tabular}
		}
	\end{table}

	After rigorous testing and evaluation, the feature set that emerged as superior, offering the most 
	consistent and highest accuracy, was F11. This feature set encompasses: (1) logarithmic 
	transformations of elemental species exospheric densities, which provide a normalized scale for 
	comparing densities of various elements; (2) the logarithm of the altitude at which measurements 
	were taken, introducing a scale that accommodates the wide range of altitudes without skewing the 
	data; (3) Sun incidence angle, accounting for the variation in solar energy impacting the elemental 
	composition; (4) the presence of H+ ions arriving through open field lines, a feature indicating solar 
	wind interaction with the planetary surface; (5) Cosine of solar time longitude, offering a representation 
	of the position in solar time longitude; and (6) Sine of latitude, providing a function to capture latitudinal 
	variations.

	The selection of F11 as the final feature set was predicated on its ability to yield the most reliable 
	and accurate predictions, thereby encapsulating the intricate dynamics and characteristics vital for 
	elemental composition analysis. This feature set's efficacy highlights the nuanced approach required 
	in feature selection to enhance neural network performance for specific predictive tasks. All the future
	training and testing were performed with this feature set as inputs to the DNN.

\subparagraph{Hyperparameter Optimization and DNN Structural Components Finalization}\hfill

	In our quest to fine-tune the multilayer perceptron for optimal performance, a significant 
	focus was placed on hyperparameter optimization. This process was critically informed by 
	the parameters outlined in Section 2.2, employing a Bayesian search strategy to navigate 
	the hyperparameter space efficiently. Our methodology involved running the optimization 
	process five times, with each iteration spanning 50 cycles and starting from a point incrementally 
	closer to the previously identified minimum, for a total of 250 cycles. This approach was instrumental 
	in inching towards the optimal hyperparameter settings, with subsequent iterations yielding diminishing returns, 
	indicative of approaching a plateau near the optimal values in the hyperparameter space. 

	During this campaign, the selection of loss functions emerged as a 
	pivotal consideration, with our experiments revealing substantial variations in model performance 
	across different functions. The discerning application of loss functions, particularly the adoption 
	of the KL-divergence for evaluating probability-like outputs, marked a jump in performance.

	The culmination of our hyperparameter optimization efforts led to the finalization of the MLP DNN 
	architecture, characterized by a four-layered structure with 600, 500, 350, and 250 neural units 
	respectively (Figure \ref{fig:mlp_final}). An adjustment was made to the regularization coefficient, setting it to the found 
	higher value of $1.0\times10^{-5}$ to enhance model generalization. Concurrently, the learning 
	rate was optimized to $0.5\times10^{-4}$, balancing the trade-off between learning speed and stability. 
	Training was conducted in mini-batches of 512 examples, a size determined through our optimization 
	exercises to be close to optimal. This meticulously optimized structure and parameter set represent 
	the culmination of our comprehensive campaign to refine the MLP DNN, ensuring it stands as a robust 
	model for our advanced predictive task.

\subparagraph{Augmented Data Study}\hfill

	In the concluding phase of our training campaign, we embarked on a strategic initiative 
	to enhance the representability of the empirical distribution, thereby aligning it more closely 
	with the true data-generating distribution—a target that remains inherently elusive due to 
	limited direct access. This endeavor was pursued through the deliberate augmentation of our 
	training datasets, an approach that involved the integration of additional examples derived 
	from the same exospheric observations that constituted our initial datasets. However, these 
	new inclusions were distinct in their representation of varying altitudes, thereby enriching the 
	diversity and depth of our training data.

	The initial expansion of our dataset to encompass 200 augmented subsets had already 
	demonstrated significant promise in enhancing the model's performance. Motivated by these 
	preliminary successes, we ambitively expanded our dataset even further to include a total of 
	300 augmented subsets, culminating in an impressive 1,360,800 examples. This substantial 
	augmentation effort was driven by the rationale that incorporating measurements from varying 
	altitudes would not only bolster the dataset's comprehensiveness but also empower our model 
	to predict with greater accuracy across a diverse range of altitude-specific inputs.

	The outcome of this labor were unmistakably positive, with the augmented datasets markedly 
	improving the robustness and accuracy of our MLP DNN, increasing further the validation set's ES4 to
	84.0\% (+1.5\%) and its R$^2$ to 63.5\% (+3.5\%). The strategic inclusion 
	of altitude-varied examples was particularly impactful, enabling the algorithm to achieve enhanced 
	predictive precision for inputs across different altitudes.

\subparagraph{Implications and Results of the Training}\hfill

	As our meticulous exploration of the hyperparameter space culminated in identifying a region that, 
	while not conclusively the ultimate minimum, demonstrates unparalleled accuracy in inferences on 
	the hold-out validation dataset, we arrived at several pivotal implications and results from our training 
	campaign. This journey through hyperparameter optimization has yielded a collection of finely tuned 
	multilayer perceptron deep neural networks, each reflecting a nuanced understanding of the underlying 
	data-generating processes.

	Firstly, one of the outcomes of this campaign is the demonstration of the algorithm's efficiency, achieving 
	optimal training within 40 complete epochs. This not only highlights the effectiveness of our chosen 
	architecture but also underscores the potential for accuracy improvements with the expansion of the 
	training dataset. Such findings affirm the architectural decisions made in designing our MLP DNN for the 
	task at hand.

	Secondly, our investigation revealed the critical role of specific features in guiding the algorithm toward 
	more precise predictions of exospheric measurements and surface composition. The identification of these 
	key features underscores the importance of thoughtful feature selection in enhancing model performance.

	Thirdly, the exploration led to the refinement of the MLP's internal structure, significantly bolstered by 
	experiments with various loss functions and output layers, alongside the application of Bayesian 
	hyperparameter optimization. While acknowledging that the realm of possible architectural enhancements 
	remains vast, the current configuration stands as a testament to the robustness and efficacy of our model.

	Lastly, the strategic augmentation of our dataset with additional exospheric observations has unequivocally 
	improved the algorithm's predictive capabilities. This expansion not only enriches the model's training 
	environment but also enhances its ability to generalize across a broader spectrum of the empirical distribution, 
	thereby moving closer to the elusive true data-generating distribution.

	The combined efforts of hyperparameter exploration, architectural fine-tuning, and dataset 
	augmentation have significantly propelled our model's performance. Through this comprehensive training 
	campaign, we have not only achieved a high degree of accuracy in our predictions but also laid a solid 
	foundation for future research to build upon, promising even greater advancements in our understanding 
	and representation of complex data-generating processes. Through this concerted effort, we have 
	significantly advanced the model's capacity to generalize from the empirical distribution to the 
	true underlying data-generating distribution.

\subparagraph{Computational Costs}\hfill

	Generating large synthetic datasets and training deep neural networks can be computationally demanding in many instances. 
	For our study, a single run of the data generation model that produces the complete exosphere above a given surface requires 
	approximately 12 minutes on a user-grade laptop equipped with a 16-core CPU and 32 GB of RAM. To produce the complete dataset 
	used for training and testing, this amounts to about 72 hours of continuous simulation and data generation time.

	Training the MLP on the largest training dataset, consisting of 1,360,800 examples, with optimized hyperparameters and a carefully 
	selected number of training epochs, takes approximately 80 minutes on the same machine. While, final model inference—predicting 
	surface composition from exospheric data—is relatively swift, typically occurring on the order of milliseconds per sample. 

	While computationally intensive during the data generation and training phases, there is still the possibility to transfer these 
	phases to supercomputers, which could significantly reduce runtime and enable the generation of even larger datasets or the training 
	of more complex models. Such an approach could also facilitate scaling the framework for broader studies, including simulations with finer 
	spatial resolution, larger parameter spaces, or higher-fidelity physical representations. Leveraging high-performance computing resources 
	would thus expand the applicability and robustness of this methodology.

\clearpage
\section{Box Plots and Map Reconstructions of the Main Test Campaign Predictions by Species}

\begin{figure}[h!]
	\centering
	\includegraphics[width=\linewidth]{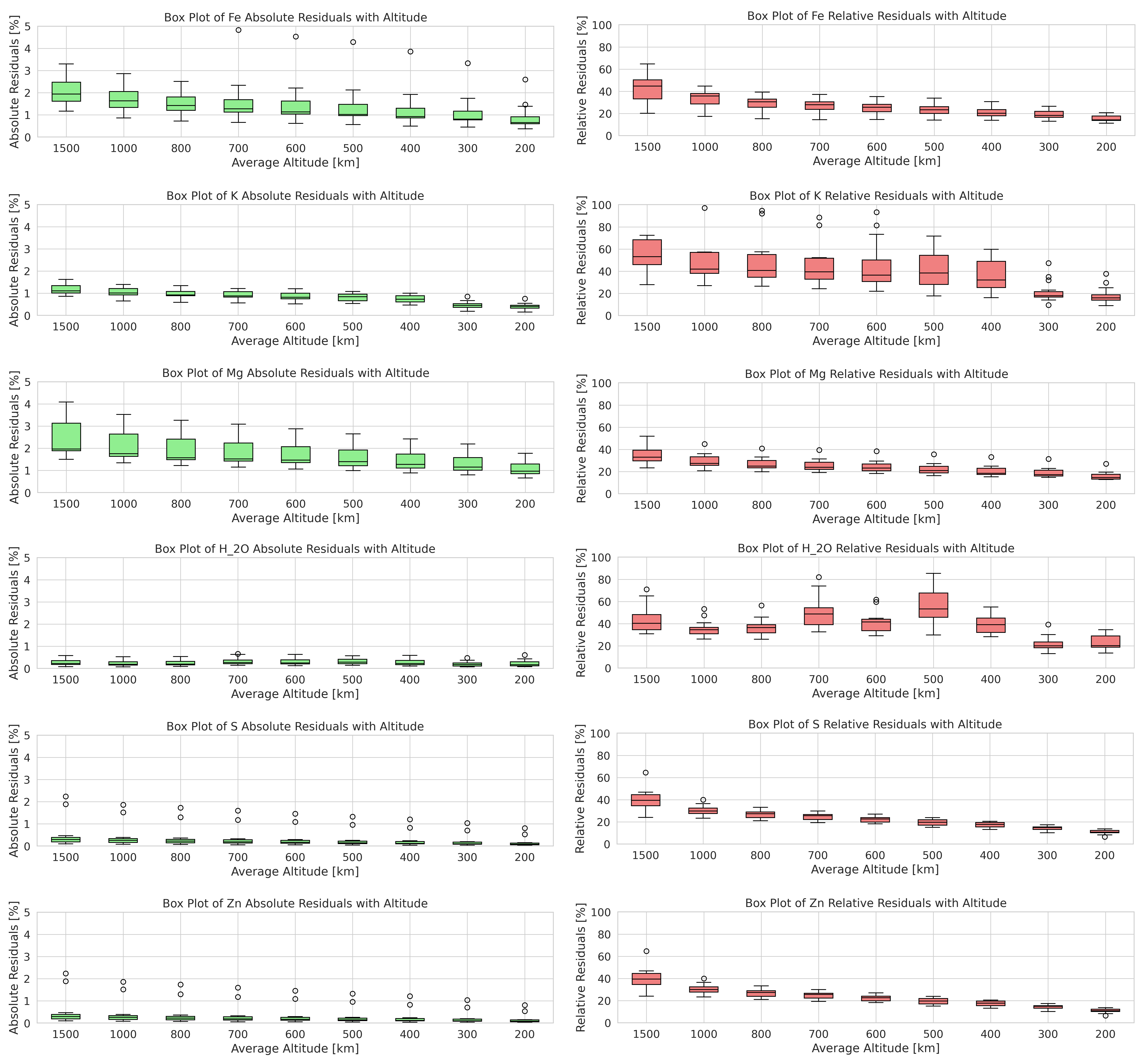}
	\caption{Box plots of the residuals metrics of the MLP DNN predictions on the 15 test surfaces of the main test 
	campaign for the element Iron, Potassium, Magnesium, Water, Sulfur, and Zinc. 
	}
  			\label{fig:boxres_fekmgh2oszn}
\end{figure}

\begin{figure}[h!]
	\includegraphics[width=\linewidth]{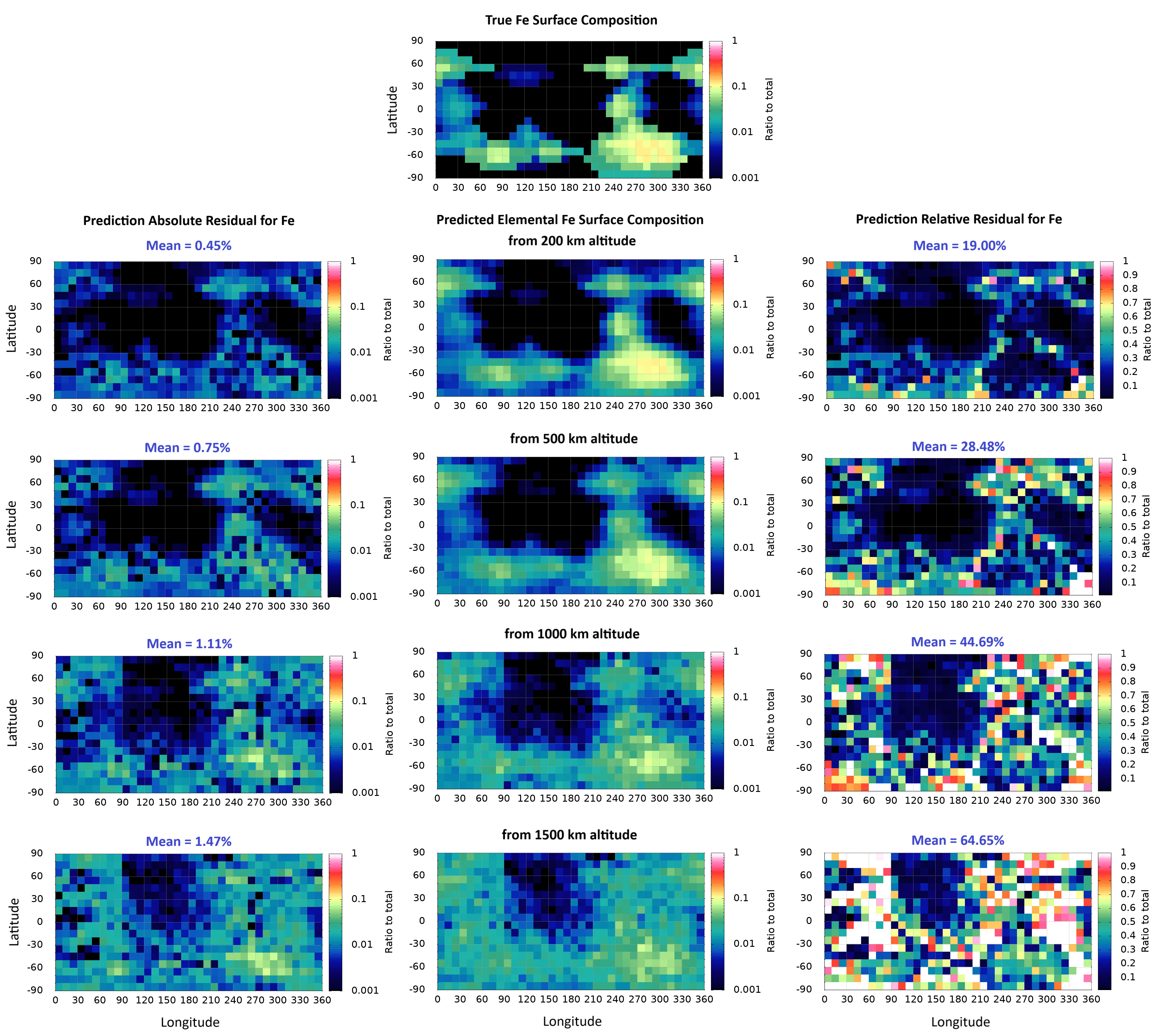}
	\caption{
	Main test campaign - MLP DNN map reconstructions of the same sample Iron 
	surface composition (test set number 2). Dayside only inputs of two simulated exospheres
	from consecutive Mercury perihelia at different altitude levels (200, 500, 1000, and 1500 km).  The top-most
	map shows the "ground truth" surface composition. The maps in the middle below it are the predicted
	fractions for this element. The panels on the left show the absolute residuals to the "ground truth", while
	on the right are the relative residuals.
	}
  	\label{fig:c_map_recon_fe}
\end{figure}

\begin{figure}[h!]
	\includegraphics[width=\linewidth]{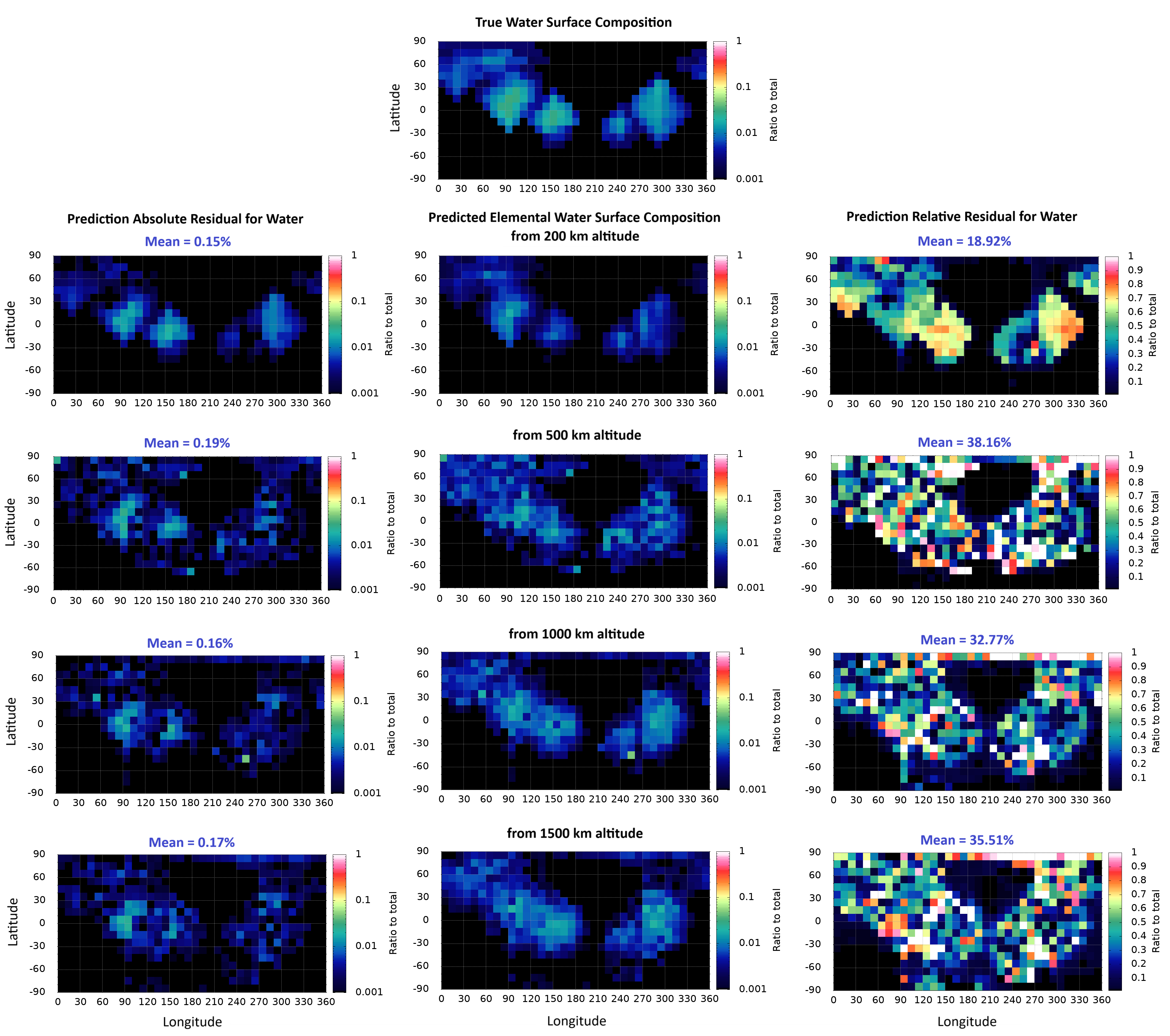}
	\caption{
	Main test campaign - MLP DNN map reconstructions of the same sample Water 
	surface composition (test set number 2). Dayside only inputs of two simulated exospheres
	from consecutive Mercury perihelia at different altitude levels (200, 500, 1000, and 1500 km).  The top-most
	map shows the "ground truth" surface composition. The maps in the middle below it are the predicted
	fractions for this element. The panels on the left show the absolute residuals to the "ground truth", while
	on the right are the relative residuals.
	}
  	\label{fig:c_map_recon_h_2o}
\end{figure}

\begin{figure}[h!]
	\includegraphics[width=\linewidth]{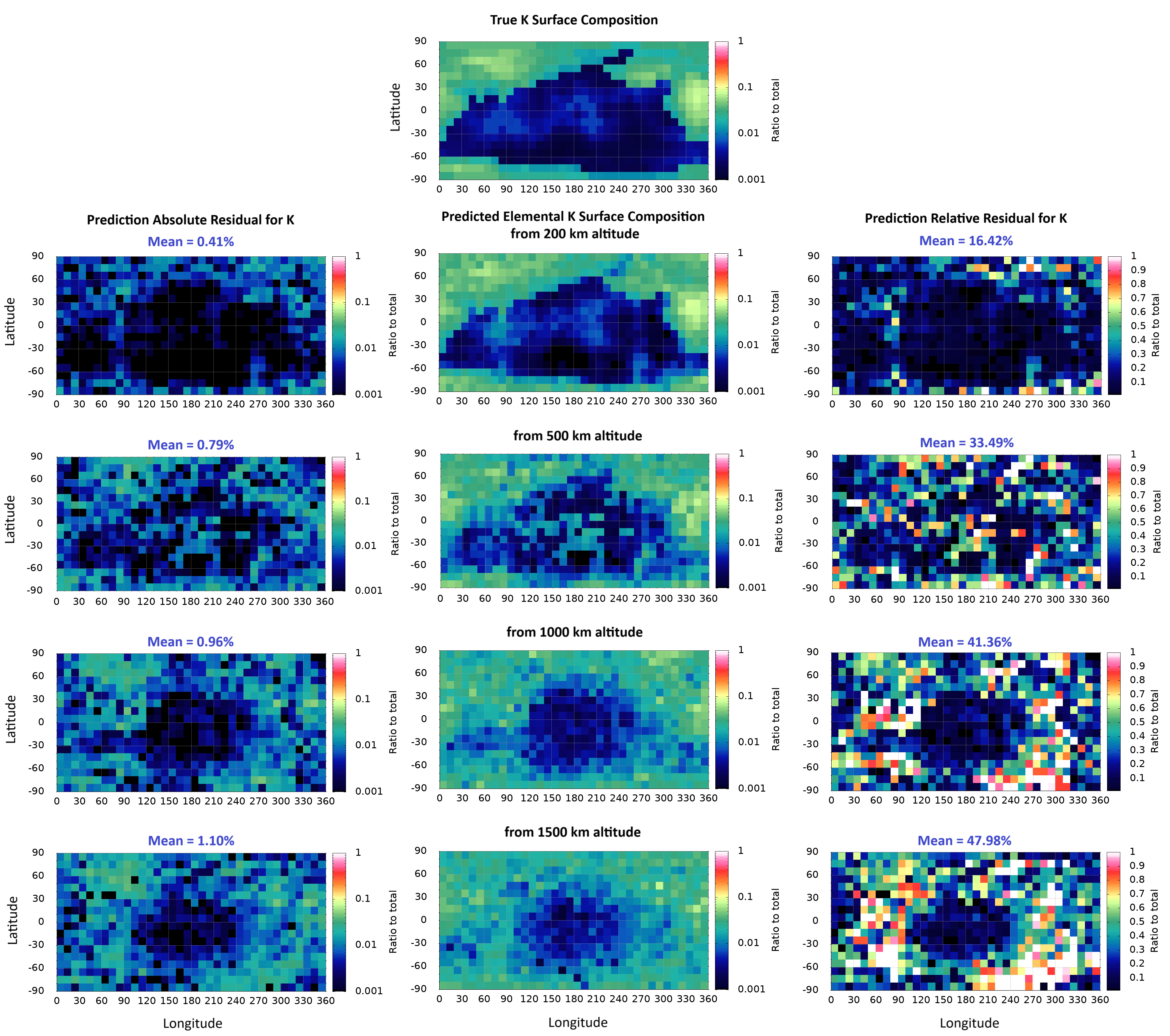}
	\caption{
	Main test campaign - MLP DNN map reconstructions of the same sample Potassium 
	surface composition (test set number 2). Dayside only inputs of two simulated exospheres
	from consecutive Mercury perihelia at different altitude levels (200, 500, 1000, and 1500 km).  The top-most
	map shows the "ground truth" surface composition. The maps in the middle below it are the predicted
	fractions for this element. The panels on the left show the absolute residuals to the "ground truth", while
	on the right are the relative residuals.
	}
  	\label{fig:c_map_recon_h}
\end{figure}

\begin{figure}[h!]
	\includegraphics[width=\linewidth]{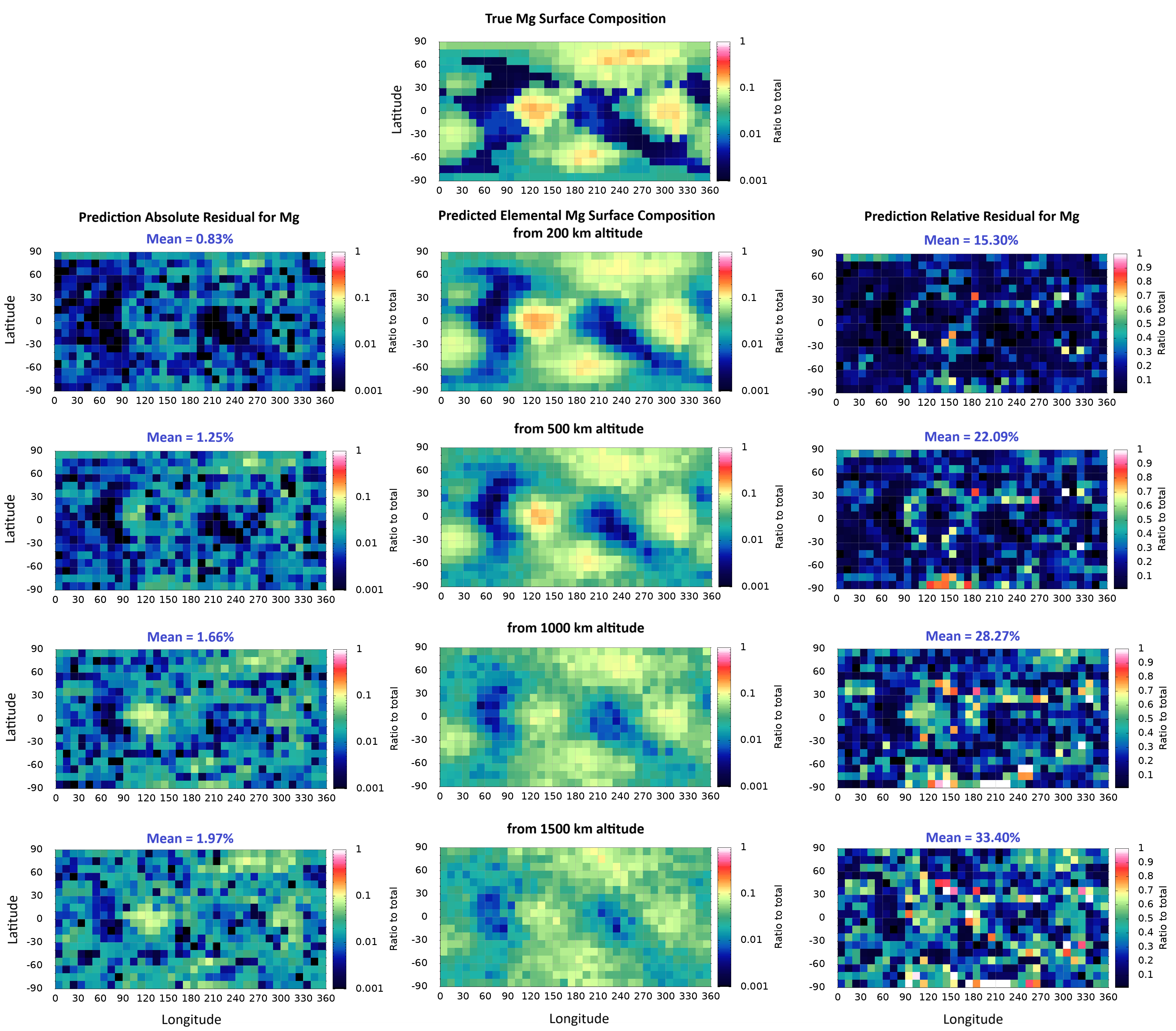}
	\caption{
	Main test campaign - MLP DNN map reconstructions of the same sample Magnesium 
	surface composition (test set number 2). Dayside only inputs of two simulated exospheres
	from consecutive Mercury perihelia at different altitude levels (200, 500, 1000, and 1500 km).  The top-most
	map shows the "ground truth" surface composition. The maps in the middle below it are the predicted
	fractions for this element. The panels on the left show the absolute residuals to the "ground truth", while
	on the right are the relative residuals.
	}
  	\label{fig:c_map_recon_mg}
\end{figure}

\begin{figure}[h!]
	\includegraphics[width=\linewidth]{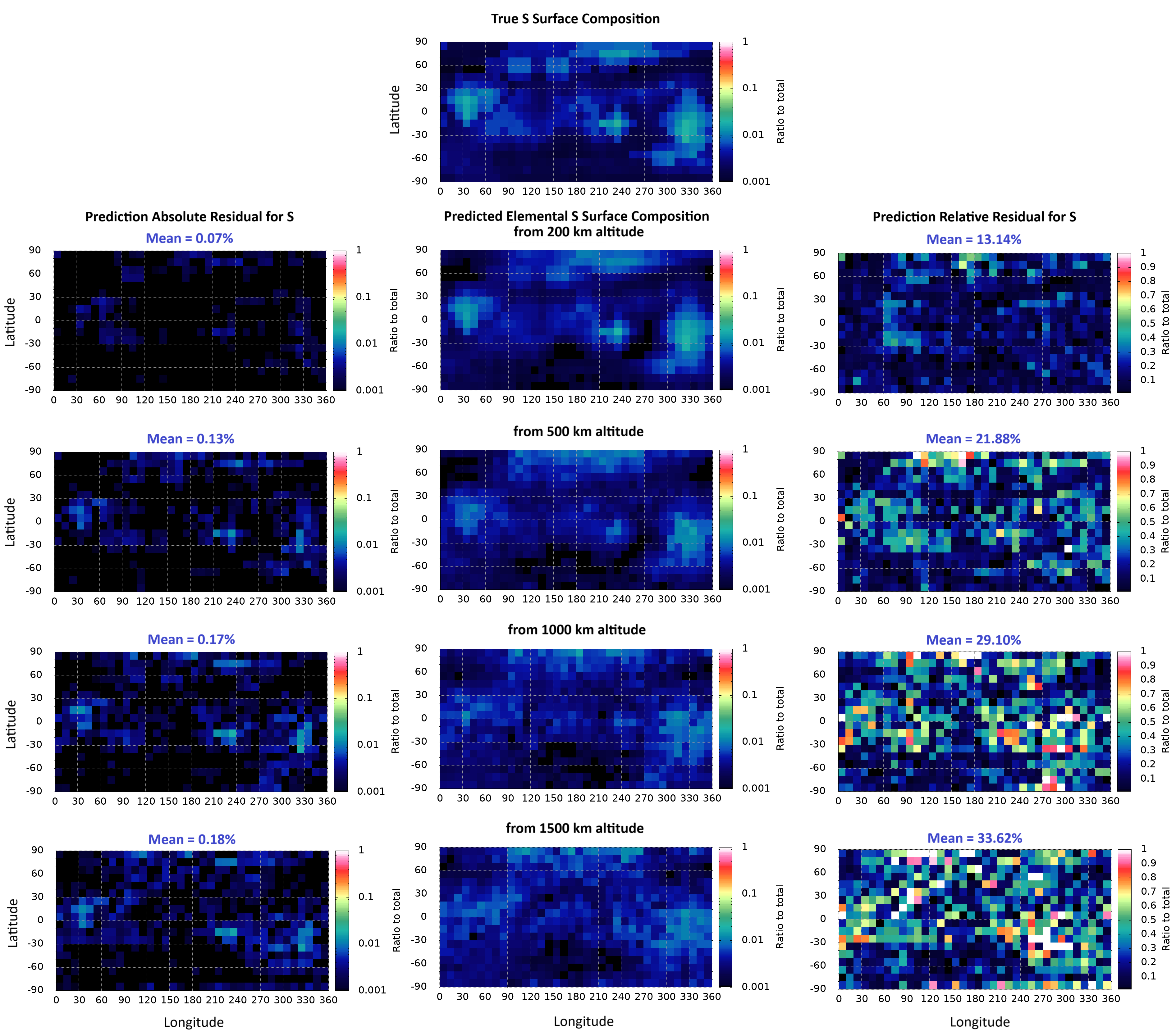}
	\caption{
	Main test campaign - MLP DNN map reconstructions of the same sample Sulfur 
	surface composition (test set number 2). Dayside only inputs of two simulated exospheres
	from consecutive Mercury perihelia at different altitude levels (200, 500, 1000, and 1500 km).  The top-most
	map shows the "ground truth" surface composition. The maps in the middle below it are the predicted
	fractions for this element. The panels on the left show the absolute residuals to the "ground truth", while
	on the right are the relative residuals.
	}
  	\label{fig:c_map_recon_s}
\end{figure}

\begin{figure}[h!]
	\includegraphics[width=\linewidth]{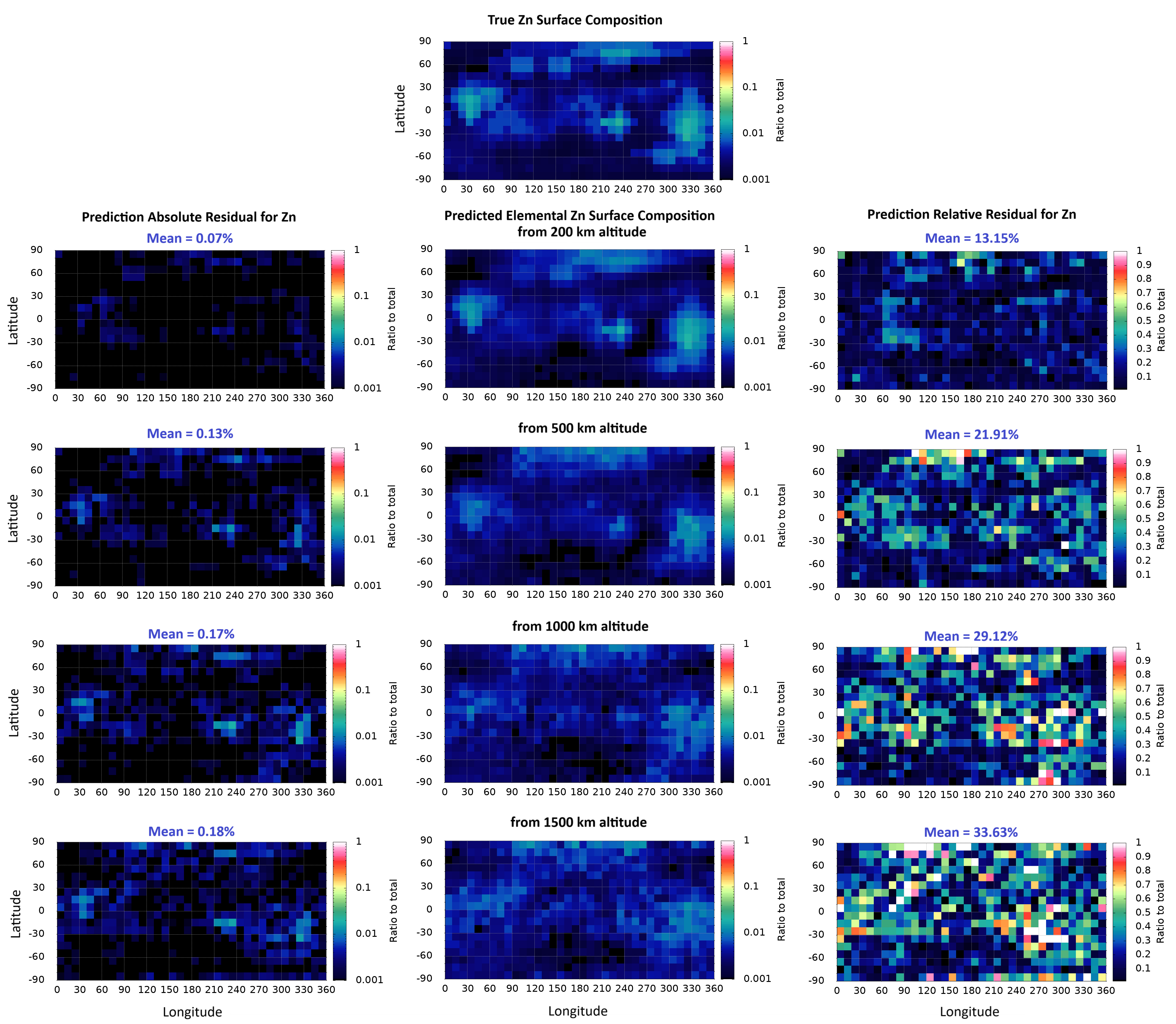}
	\caption{
	Main test campaign - MLP DNN map reconstructions of the same sample Zinc 
	surface composition (test set number 2). Dayside only inputs of two simulated exospheres
	from consecutive Mercury perihelia at different altitude levels (200, 500, 1000, and 1500 km).  The top-most
	map shows the "ground truth" surface composition. The maps in the middle below it are the predicted
	fractions for this element. The panels on the left show the absolute residuals to the "ground truth", while
	on the right are the relative residuals. Notably, the Zinc reconstructed maps correctly follow the Sulfur ones due
	to the matching source "mineral" for the two elements in our setting.
	}
  	\label{fig:c_map_recon_zn}
\end{figure}

\begin{figure}[h!]
	\includegraphics[scale=0.16]{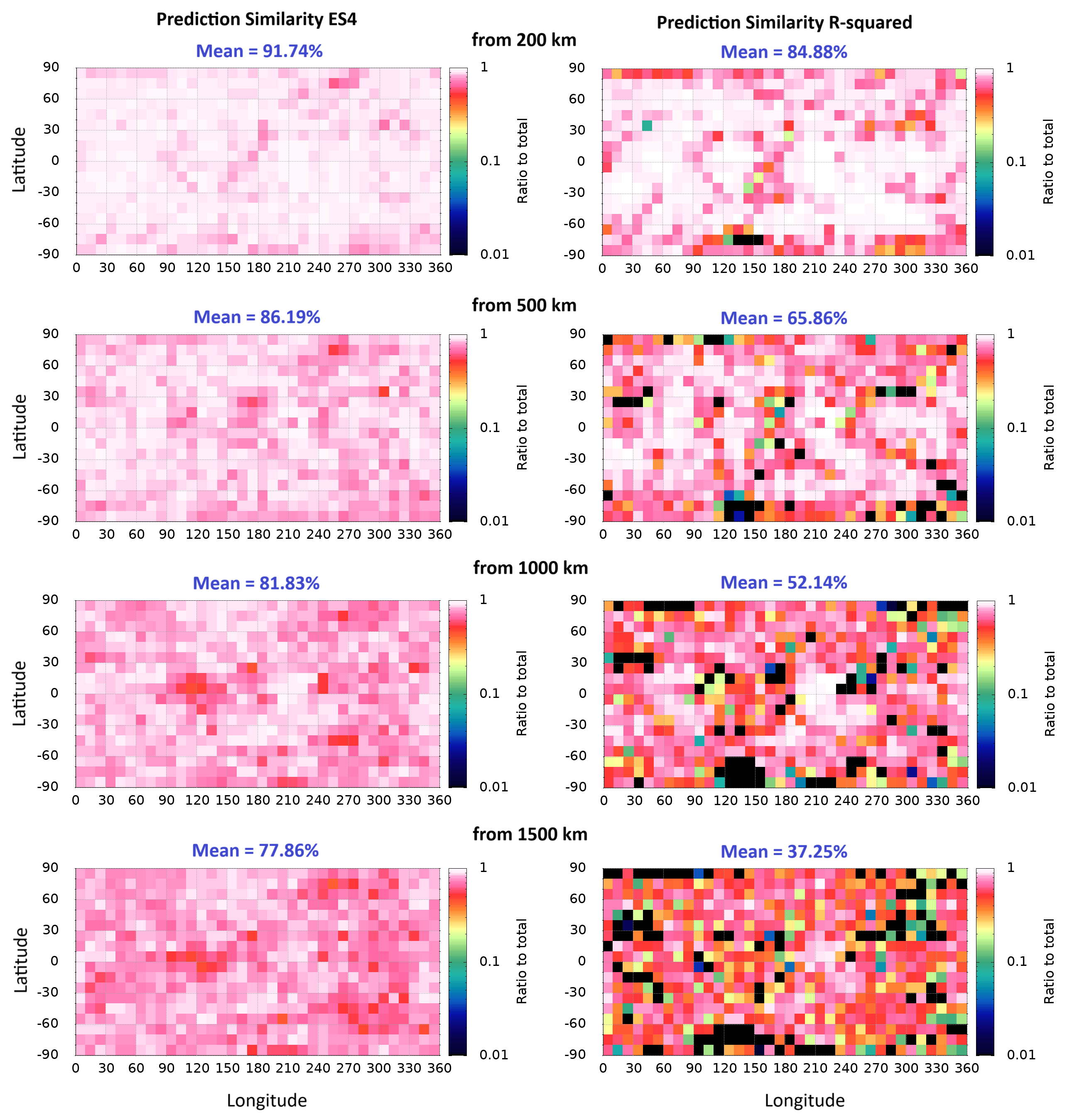}
	\caption{
	Map representations of the accuracy metrics evaluated for the predictions by the finalized trained MLP DNN 
	of the same sample total surface composition (test set number 2). Dayside only inputs of two 
	simulated exospheres from two consecutive Mercury perihelia at different altitude levels (200, 500, 1000, and 
	1500 km). The panels on the left show the ES4 similarity, while on the right are the R-squared metrics.
	}
  	\label{fig:c_map_recon_metrics}
\end{figure}

\clearpage

\section*{Open Research Section}

The data used for training and testing the DNN of the main testing campaign (both inputs and outputs) 
are available at Zenodo via DOI 10.5281/zenodo.13778487
(URL: \url{https://zenodo.org/records/13778487}) with a Creative Commons Attribution license \cite{kazakov_io_dnn}.
The inputs and outputs data to the exosphere generation model are available at Zenodo via DOI 10.5281/zenodo.13780529 (URL:  \url{https://zenodo.org/records/13780529}) 
with Creative Commons Attribution license \cite{kazakov_io_exo}.
The data for all the various surfaces used to produce the exospheres and to train and test the algorithm are
available at Zenodo via DOI 10.5281/zenodo.13780740 (URL:  \url{https://zenodo.org/records/13780740}) with Creative Commons Attribution license \cite{kazakov_surf}.

Version 0.1.0 of the PTF-A-MLP software used to train and test the deep neural network is available at GitHub-Zenodo 
via doi 10.5281/zenodo.13785015 (URL:  \url{https://zenodo.org/records/13785015}) with Creative Commons Attribution license \cite{Kazakov_PTF-A-MLP_2024} and
developed openly at  \url{https://github.com/AJKazakov/PTF-A-MLP}. The software is written in Python, and a simple 
usage instruction is included in the README file. Please direct any requests for clarification on its usage to the first author.

The exospheric model used to generate the exosphere is available at:  \url{http://150.146.134.250/cgi-bin/modello-input.pl} and is part
of the SPIDER (Sun Planet Interactions Digital Environment on Request) interface, accessible at \url{http://spider-europlanet.irap.omp.eu/}.

\acknowledgments
This work is supported by the Italian Space Agency (ASI) and by the Italian National Institute of Astrophysics (INAF): 
SERENA agreement no. 024-66-HH.0 “Attività scientifiche per il Payload SERENA su BepiColombo, relative alla fine 
della fase di crociera e fase operativa”.

\bibliography{Kazakov_et_al_2025_arxiv.bib}

\end{document}